\documentclass{article}

\usepackage{amssymb}
\usepackage{graphicx}
\usepackage{epsfig}

\def\bb {\begin {eqnarray}}
\def\ee {\end {eqnarray}}

\begin{document}
\begin{center}
{\Large 
Adiabatic  description of nonspherical quantum dot models
\footnote{Submitted to Physics of Atomic Nuclei}} \\ [5mm]

{\large \it  A.A. Gusev \footnote{e-mail: gooseff\@jinr.ru}, O. Chuluunbaatar, S.I. Vinitsky}\\ [5mm]

Joint Institute for Nuclear Research, Dubna, Russia\\ [5mm]

{\large \it K.G. Dvoyan, E.M. Kazaryan, H.A. Sarkisyan}\\ [5mm] 
 
 Russian-Armenian (Slavonic) University, Yerevan, Armenia\\ [5mm]

{\large V.L. Derbov, A.S. Klombotskaya, V.V. Serov}\\ [5mm]

Saratov State University, Saratov, Russia\\ [5mm]

\textbf{Abstract}
\end{center}

Within the effective mass approximation an \textit{adiabatic} description
of  spheroidal and dumbbell quantum dot models in the regime of strong dimensional quantization
is presented using the expansion of the wave function in appropriate
sets of single-parameter basis functions.
The comparison is given and  the peculiarities are considered for
spectral and optical characteristics of the models with
axially symmetric confining potentials depending on their geometric size
making use of the total sets of exact and \textit{adiabatic} quantum numbers
in appropriate analytic approximations.

\textbf{Key words:}
 spheroidal and dumbbell quantum dot models, boundary-value problem, Kantorovich method, adiabatic approximation, absorption coefficient 

\centerline{Blessed memory of
professor Sissakian Alexei Norairovich is devoted}

\section{Introduction}

To analyze the geometrical, spectral and optical characteristics of
quantum dots in the effective mass approximation and in the regime
of strong dimensional quantization following ~\cite{Harrison}, many
methods and models were used.
{ We mention some of them, that are in the field of our
interest:}  the exactly solvable models of spherical
and cylindrical layer (toroid)
impermeable wells ~\cite{Gambaryan,Hayrutyunyam},
the adiabatic approximation for a lens-shaped
well confined to a narrow wetting layer~\cite{Hawrylak96},
and a
hemispherical impermeable well~\cite{Hayk},
the model of strongly
oblate or prolate ellipsoidal  impermeable well ~\cite{79,79a,KimZubarev}, as
well as numerical solutions of the boundary value problems (BVPs)
with separable variables in the spheroidal coordinates for wells
with infinite and finite wall heights
~\cite{CNI2000,Trani,Lepadatu,spherqd,Filikhin}, M\"obius~\cite{mobius} nanostructures,
scattering problems for toric ~\cite{afanasev01}
and coupled nonidentical microdisks ~\cite{avoid}.

Similar models were used for describing the energy spectra of deformed
nuclei~\cite{GS51,Rassey58,Arvieu87,Arvieu93,Pashkevich69,P69ashkevich,Greiner72},
atomic clusters deposited on planar surfaces \cite{Poenaru08}
and
low-energy barrier nuclear
reactions~\cite{Hofmann74,Furman,cpc99,Zagrebaev04,Zagrebaev07}. However,
thorough comparative analysis of spectral and { optical}
characteristics of models with different potentials, including those
with non-separable variables, remains to be a challenging problem.

In the present paper we analyze the spectral {and optical}
characteristics of the following models: a spherical quantum dot
(SQD), an oblate spheroidal quantum dot (OSQD), a prolate spheroidal
quantum dot (PSQD), and a dumbbell QDs (DQD). We make use of the
Kantorovich method that reduces the problem to a set of ordinary
differential equations (ODE)~\cite{put}
by means of the expansion of the wave function in appropriate sets
of single-parameter basis functions~\cite{CADE09} similar to the
well-known adiabatic method~\cite{BhK58}.

We present briefly a calculation scheme for solving elliptical BVPs
with axially-symmetric potentials in cylindrical coordinates (CC),
spherical coordinates (SC), oblate spheroidal coordinates (OSC), and
prolate spheroidal coordinates (PSC). Basing on the
symbolic-numerical algorithms (SNA) developed for axially-symmetric
potentials~\cite{CASC07,CASC09,CASC10}, different sets of
solutions are constructed for the parametric BVPs related to the
fast subsystem, namely, the eigenvalue problem solutions (the
terms and the basis functions), depending upon the slow variable
as a parameter, as well as the matrix elements, i.e., the
integrals of the products of basis functions and their derivatives
with respect to the parameter. These terms and matrix elements
form the matrices of variable coefficients in the set of
second-order ODE with respect to the slow variable, which are
calculated in special cases analytically and in the general case
using the program ODPEVP~\cite{ODPEVP}. The BVP for this set of
ODEs is solved by means of the program KANTBP~\cite{kantbp}, while
in the special cases crude diagonal estimations can be performed
using the appropriate analytic approximations.

The efficiency of the calculation scheme and the SNA used is
demonstrated by tracing the peculiarities of spectral and optical
characteristics in the course of varying the  ellipticity of the
prolate or oblate spheroid and dumbbell in the models of quantum
dots with different confining potentials, such as the isotropic
and anisotropic harmonic oscillator, the spherical and spheroidal
well with finite or infinite walls approximated by smooth
short-range potentials, as well as by constructing the adiabatic
classification of the states.

The paper is organized as follows. In Section \ref{1}, the
calculation scheme for solving elliptic BVPs with
axially-symmetric confining potentials is briefly presented.
Sections \ref{3}  and \ref{5} are devoted to the analysis of the
spectra { and absorption coefficient} of quantum dot models with
three types of axially-symmetric potentials, including the benchmark
exactly solvable models. In Conclusion we
summarize the results and discuss the future applications.

\section{Problem Statement}\label{1}
Within the effective mass approximation under the conditions of
strong dimensional quantization, the Schr\"{o}dinger equation for
the slow envelope of the wave function $\tilde\Psi(\tilde{\vec r})$
of a charge carrier (electron $e$ or hole $h$) in the models of
QDs has the form~\cite{79,79a}
\begin{eqnarray}
\label{sp01}
 \{\tilde {\hat H} -\tilde E \}\tilde \Psi(\tilde {\vec r}) =
 \{
({2\mu_p })^{-1} \tilde {\hat { {\vec P}}}^2 + \tilde U ( \tilde
{\vec r}) -\tilde E
 \}\tilde \Psi(\tilde {\vec r}) =0,
\end{eqnarray}
where $\tilde {\vec r}\in \bf R^{3}$ is the position vector of the
particle having the effective mass $\mu_{p}=\mu_{e}$ (or
$\mu_{p}=\mu_{h}$), $\tilde {\hat {{\vec P}}}=-i\hbar \nabla
_{\tilde {\vec r}}$ is the momentum operator,
 $\tilde E$ is the energy of the particle,
 $\tilde U ( \tilde {\vec r})$ is the axially-symmetric potential confining
the particle motion in SQD, PSQD, or OSQD. In Model A,   $\tilde U(
\tilde {\vec r})$ is chosen to be the potential of an isotropic or
anisotropic axially-symmetric harmonic oscillator in Cartesian coordinates ${\vec r=\{x,y,z\}}$:
\begin{eqnarray}
\label{sp02}
\tilde U^{\mbox{A}} (\tilde {\vec r}) =
{\mu_p \tilde \omega ^2( \zeta_1(\tilde x^2+\tilde y^2)+\zeta_3\tilde z^2)}/{2}.
\end{eqnarray}
Here $\zeta_1=1$, $\zeta_3=1$ for a spherical QD
or $\zeta_1= (\tilde r_0/\tilde a)^4$,
$\zeta_3=(\tilde r_0/\tilde c)^4$  for a spheroidal QD, inscribed into a spherical
one, where $\tilde a$ and $\tilde c$ are the semiaxes of the
ellipse which transforms into a sphere at $\tilde a=\tilde
c=\tilde r_0=\sqrt{\tilde x_0^2+\tilde y_0^2+\tilde z_0^2}$,
 $\tilde \omega =\gamma _{\tilde r_0 } {\hbar }/({\mu_p
\tilde r_0^2 })$ is the angular frequency,  and $\gamma_{\tilde r_0 }$  is an
adjustable parameter. We will use the value $\gamma_{\tilde r_0 }= \pi^2/3$ that follows
from equating the ground state energies for the spherical oscillator and the
spherical QD of Model B considered below. If necessary, this definition
can be replaced with a different one, e.g., the one
conventional for nuclear physics \cite{Pashkevich69,P69ashkevich,Greiner72}.

For Model B,   $\tilde U (\tilde {\vec r})$ is the
potential of a spherical or axially-symmetric well
\begin{eqnarray}
\label{sp03}
\tilde U^{\mbox{B}} (\tilde {\vec r})
=\{0,S(\tilde {\vec r}) < 0 ; \tilde U_0,
S(\tilde {\vec r}) \ge 0 \},
\end{eqnarray}
bounded by the surface  $S(\tilde {\vec r})=0$
with  walls of finite or infinite height $1\ll \tilde U_0 <
\infty$.
In Eq. (\ref{sp03}) $S(\tilde {\vec r})$ depends on the parameters $\tilde a$, $\tilde c$,
and $0\le \tilde c_1 \le 1$
\begin{eqnarray}   \label{eq99a}
S(\tilde {\vec r})\equiv\frac{\tilde x^2+\tilde y^2}{\tilde a^2}
+\frac{(\tilde z^2-\tilde c^2)(\tilde z^2\tilde c_1^2+1-\tilde c_1^2)^2}
{\tilde c^2(\tilde c_1^2\tilde c^2/4+1-\tilde c_1^2)^2}
\end{eqnarray}
At $c_1=0$ we get a spheroidal quantum dot model, at $0<c_1<1$
it becomes a dumbbell QD with a symmetric double well,
and at  $c_1>1$ we get a triple-well model.

For Model C,   $\tilde U ( \tilde {\vec r})$  is taken to be a
spherical or axially-symmetric diffuse potential
\begin{eqnarray}
\label{sp03VS}
\tilde U^{\mbox{C}} (\tilde {\vec r})
= \tilde U_0\left(1-\left(1+\exp (S(\tilde {\vec r})/{s})\right)^{-1}\right),
\end{eqnarray}
where $s$ is the edge diffusiveness parameter of the function
smoothly approximating the vertical walls of finite height
$\tilde U_0$. Below we restrict ourselves by considering Model B
with infinite walls $\tilde U_0 \rightarrow\infty$ and Model C
with walls of finite height $\tilde U_0$.

Throughout the paper we make use of the reduced atomic
units~\cite{Harrison,79a}: $a_B^*= {\kappa \hbar ^2}/{\mu _p e^2}$
is the reduced Bohr radius, $\kappa$ is the DC permittivity,
 $\tilde E_R \equiv Ry^*= {\hbar ^2}/({2\mu _p {a_B^*}^2 })$
is the reduced Rydberg unit of energy, and the following
dimensionless quantities are introduced: $\tilde \Psi(\tilde{\vec
r})= {a_B^*}^{-3/2}\Psi( {\vec r})$, $2 {\hat H}= \tilde {\hat
H}/{Ry^*}$,
$2 {E}= {\tilde E}/{Ry^*}$, $2 {U(\vec r)}= {\tilde U (\tilde {\vec
r})}/{Ry^*}$, $\vec r=\tilde {\vec r} /a_B^*$, $a=\tilde a/a_B^*$,
$\tilde c=c/a_B^*$,$\tilde c_{1}=c_{1}/a_B^*$, $r_0=\tilde
r_0/a_B^*$, $\omega=\gamma_{r_{0}}/ r_0^{2}=\hbar
\tilde\omega/(2Ry^*)$. For an electron with the effective mass
$\mu_p \equiv \mu_e = 0.067 m_0 $ at $\kappa =13.18$ in GaAs:
$a_B^*=a_{B}^{e}=104\;{}^{{}^o}\!\!\!\!A =10.4$ nm and $Ry^*=\tilde E^{e}_{R}
=5.275$ meV.
For a heavy hole with the effective mass
$\mu_{h}=\mu_{e}/0.12=0.558m_{0}$ the corresponding values are
$a_{B}^{h}=a_{B}^{e}( \mu_{e}/\mu_{h})= 12.48 \;{}^{{}^o}\!\!\!\!A =1.248$ nm, and
$\tilde E^{h}_{R}=\tilde E^{e}_{R}( \mu_{h}/\mu_{e})=46.14 $ meV.

Note, that for model A of approximation of OSQD/PSQD by the anisotropic oscillator
(\ref{sp02}) the separation of variables
in cylindric coordinates $\vec x=(z,\rho,\varphi)$ is possible and additional integrals exist
\cite{Demkov1,Demkov,Ilkaeva}.
Similarly, for model B the variables are separable in the oblate/prolate
spheroidal coordinates $\vec x=(\xi,\eta,\varphi)$ and the
additional integrals of motion are
$\hat \Lambda$: $[\hat H,\hat \Lambda]\equiv \hat H\hat \Lambda-\hat
\Lambda\hat H=0$, i.e. $\hat H_p$ and  $\hat \Lambda_p$ in PSQD
{\footnotesize
\begin{equation}\label{la1948o}
\hat H_p=-\frac{4}{d^2}\left[\frac{1}{\xi^2-\eta^2}
\left(\frac{d}{d\xi}(\xi^2-1)\frac{d}{d\xi}
+\frac{d}{d\eta}(1-\eta^2)\frac{d}{d\eta}\right)
 +\biggl(\frac{1}{(\xi^2-1)(1-\eta^2)}\biggr)\frac{d^2}{d\varphi^2}\right],
\end{equation}
\begin{equation}\label{la1948}
\hat
\Lambda_p=\frac{1-\eta^2}{\xi^2-\eta^2}\frac{d}{d\xi}(\xi^2-1)\frac{d}{d\xi}
+\frac{\xi^2-1}{\xi^2-\eta^2}\frac{d}{d\eta}(1-\eta^2)\frac{d}{d\eta}
 +\biggl(\frac{1}{\xi^2-1}-\frac{1}{1-\eta^2}\biggr)\frac{d^2}{d\varphi^2},
\end{equation}
}   $\hat H_o$ and  $\hat \Lambda_o$ in OSQD
{\footnotesize \begin{equation}\label{la1949o}
\hat
H_o=-\frac{4}{d^2}\left[\frac{1}{\xi^2+\eta^2}\left(\frac{d}{d\xi}(\xi^2+1)\frac{d}{d\xi}
+\frac{d}{d\eta}(1-\eta^2)\frac{d}{d\eta}\right)
 -\biggl(\frac{1}{(\xi^2+1)(1-\eta^2)}\biggr)\frac{d^2}{d\varphi^2}\right],
\end{equation}
\begin{equation}\label{la1949}
\hat
\Lambda_o=-\frac{1-\eta^2}{\xi^2+\eta^2}\frac{d}{d\xi}(\xi^2+1)\frac{d}{d\xi}
-\frac{\xi^2+1}{\xi^2+\eta^2}\frac{d}{d\eta}(1-\eta^2)\frac{d}{d\eta}
 -\biggl(\frac{1}{\xi^2+1}+\frac{1}{1-\eta^2}\biggr)\frac{d^2}{d\varphi^2}.
\end{equation}}
Eq. (\ref{la1949}) is obtained by substituting $\xi
\rightarrow \imath \xi$, $d\rightarrow -\imath d$ from the known
(\ref{la1948}) derived in \cite{Erikson,MPST85}.

Since the Hamiltonian $\hat H$ in Eqs. (\ref{sp01})--(\ref{sp03VS}) commutes with the $z$-parity operator of reflection in the plane $z=0$
($z \to - z$  or $\eta  \to -\eta $), the solutions are divided into
even ($\sigma = + 1$) and odd ($\sigma = - 1$) ones.
The solution of Eq. (\ref{sp01}),  periodical with respect to the azimuthal angle
$\varphi$,  is sought in the form of a product
$\Psi(x_{f},x_{s},\varphi) = \Psi^{m\sigma} (x_{f},x_{s}
){e^{im\varphi }}/{\sqrt {2\pi } }$, where $m = 0,\pm 1,\pm 2,...$
 is the magnetic quantum number. Note, in absence of magnetic fields the Hamiltonian commutes also with the inversion operator ($\vec r\to -\vec r$) with eigenvalues $\hat{\sigma}=(-1)^{m}\sigma$ and solution divided into {\it gerade} ($\hat\sigma = + 1$) and {\it ungerade} ($\hat\sigma = - 1$) ones.
Then the function
$\Psi^{m\sigma} (x_{f},x_{s} )$
satisfies the following equation in the two-dimensional domain
$\Omega=\Omega_{x_f}(x_s)\cup\Omega_{x_s}\subset {\bf R}^2\backslash
\{0\}$, $\Omega_{x_f}(x_s)=( x_f^{\min }(x_{s}),x_f^{\max
}(x_{s}))$, $\Omega_{x_s}=( x_s^{\min },x_s^{\max })$:
\begin{eqnarray} &&
\left( \hat H_1 (x_{f};x_s) + \hat H_2 (x_{s})+ V (x_{f},x_{s}) - 2
{E}\right) \Psi^{m\sigma} (x_{f},x_{s} ) = 0. \label{sp09}
\end{eqnarray}
The Hamiltonian of the slow subsystem $\hat H_2(x_s)$ is expressed as
\begin{eqnarray} &&\label{sp09xs}
 \hat H_2(x_s)= \check H_2(x_s)
 = - \frac{1}{g_{1s}(x_{s})}\frac{\partial }{\partial x_{s}}g_{2s}(x_{s})
 \frac{\partial }{\partial x_{s} }+ \check V_{s}(x_{s}),
\end{eqnarray}
and the Hamiltonian of the fast subsystem
$ \hat {H}_1 (x_f ;x_s )$
is expressed in terms of the reduced Hamiltonian  $\check H_f (x_f ;x_s)$
and the weighting factor $g_{3s}(x_s)$:
\begin{eqnarray}\label{sp09xf}&&
  \hat H_1(x_{f};x_s)=g_{3s}^{-1} (x_s )\check H_f (x_f ;x_s),\\&&\nonumber
 \check H_f (x_f ;x_s)= - \frac{1}{g_{1f}(x_{f})}\frac{\partial }{\partial x_{f}}g_{2f}(x_{f})
 \frac{\partial }{\partial x_{f} }+ \check V_{f}(x_{f})
 +\check V_{fs}(x_{f},x_{s}).
\end{eqnarray}
\begin{table}[t]\caption{The values of conditionally fast $x_{f}$ and slow $x_{s}$
independent variables,
the coefficients $g_{is}(x_s)$, $g_{jf}(x_f)$
and the potentials $\check V_{f}(x_{f})$, $\check V_{s}(x_{s})$, $\check V_{fs}(x_{f},x_{s})$,
in Eqs.(\ref{sp09})--(\ref{sp09xf})
 for SQD, OSQD and PSQD in cylindrical (CC), spherical (SC) and
 oblate $\&$ prolate spheroidal ({OSC} $\&$ {PSC}) coordinates with
$(d/2)^{2}=\pm( a^2 - c^2)$, $+$ for OSC, $-$  for PSC.}\label{ktpy}
\parbox{0.68\textwidth}{\footnotesize
\begin{tabular}
{|c|c|c|c|c|} \hline & \multicolumn{2}{|c|}{CC} &
\multicolumn{1}{|c|}{SC} &{OSC} $\&${PSC} \\\hline
\hline & OSQD& PSQD& SQD & OSQD $\&$ PSQD \\
\hline $x_f $& $ z $& $ \rho$& $\eta$&$\eta$
 \\
\hline $x_s $& $ \rho $& $ z$&
$ r $&$\xi$\\
\hline $g_{1f} $&$1$ & $ \rho $&
$1$&$1$\\
\hline $g_{2f} $&$1$ & $ \rho $&
$1 - \eta ^2 $&$1 - \eta ^2 $\\
\hline $g_{1s}$ & $ \rho $ & $1$&
$ r^2 $ &$1$\\
\hline $g_{2s} $& $ \rho $& $1$&
$ r^2 $&$\xi^2\pm 1$\\
\hline $g_{3s} $& $1$& $1$&
$ r^2 $&$1$ \\
\hline ${\check V}_f(x_f)$& $ \omega^2\zeta_3 z^2$& $m^{2}/\rho^{2}+
\omega^2\zeta_1 \rho^2$& $
m^{2}/g_{2f} $&$
m^{2}/g_{2f} \pm (d/2)^{2}g_{2f}2E
$\\
\hline ${\check V}_s(x_s)$&$m^{2}/\rho^{2}+ \omega^2\zeta_1 \rho^2$
& $ \omega^2\zeta_3 z^2$& $0$&$
\mp m^{2}/g_{2s}- ((d/2)^{2}g_{2s} - 1)2E
$\\
\hline ${\check V}_{fs}(x_f,x_s)$& $0$&$0$&
$ {\check V}( r,\eta) $&$ {\check V}(\xi,\eta)$\\
\hline
\end{tabular}
}
\end{table}

Table \ref{ktpy} contains a detailed description
of the conditionally fast $x_{f}$
and slow $x_{s}$ independent variables, the coefficients
$g_{1s}(x_s)$, $g_{2s}(x_s)$, $g_{3s}(x_s)$, $g_{1f}(x_f)$,
$g_{2f}(x_f)$, and the reduced potentials $\check V_{f}(x_{f})$,
$\check V_{s}(x_{s})$, $\check V_{fs}(x_{f},x_{s})$, entering Eqs.
(\ref{sp09})--(\ref{sp09xf}) for SQD, OSQD, and PSQD in
cylindrical  ($\vec x=(z,\rho,\varphi)$), spherical
 ($\vec x=(r,\eta=\cos\theta,\varphi)$),
 and oblate/prolate spheroidal ($\vec x=(\xi,\eta, \varphi)$)
 coordinates
 (CS, SC and OSC/PSC) \cite{stigun}.
Note, that in Table \ref{ktpy}, using  Eqs. (\ref{sp02}), (\ref{sp03VS})
in the reduced atomic units, the potential $\check V(r,\eta)$ for OSQD/PSQD in SC
is expressed for Model A as
$$ \check V(r,\eta)={2}r ^2 U^A(r,\eta)
={\omega ^2r ^4( \zeta_1(1-\eta^2)+\zeta_3\eta^2)},$$
and for Model C as
$$\check V(r,\eta)={2}r ^2 U^C(r,\eta)
={2}r ^2U_0\left(1-\left(1+\exp (({  r^2 (
\frac{(1-\eta^2)}{a^2} +\frac{\eta^2}{c^2})-1})/{s})\right)^{-1}\right),$$ both
having  zero normal first derivatives $\partial V( r,\eta)/\partial
r$ in the vicinity of the origin $r=0$ (equilibrium point), similar
to \cite{BuckPilt}. We do not use the CC for Model C, because the
motion in this case is not restricted by two coordinates $\rho$ and
$z$. For Model B in Table \ref{ktpy} $\omega=0$ and the potentials
${\check V}(r,\eta)={\check V}(\xi,\eta)=0$ are zero, since in this
case one should impose the Dirichlet boundary conditions
$\Psi^{m\sigma} (x_{f},x_{s} )|_{\partial \Omega}=0$ at the boundary
$\partial \Omega=\{{\cal R}^2 | S(x_{f},x_{s})=0\}$ of $\Omega$
restricted by the surface $S(\tilde {\vec r})=0$, which is
equivalent to the action of the potential (\ref{sp03}).

The solution  $\Psi^{m\sigma}_i(
x_{f},x_{s})\equiv\Psi^{Em\sigma}_i( x_{f},x_{s})$ of the problem
(\ref{sp09})--(\ref{sp09xf}) is sought in the form of Kantorovich
expansion \cite{put}
\begin{eqnarray}
  \Psi^{Em\sigma}_i( x_{f},x_{s})=\sum_{j=1}^{j_{\max}}
 \Phi^{m\sigma}_j(x_{f}; x_{s})\chi_{j}^{(m\sigma i)}(E,x_{s}).
 \label{sp15}
\end{eqnarray}
The set of appropriate trial functions is chosen as the set of eigenfunctions
$\Phi^{m\sigma}_j(x_{f}; x_{s})$  of the Hamiltonian $\check {H}_f
(x_f ;x_s )$ from  (\ref{sp09xf}),
i.e., the solutions of the parametric BVP
\begin{equation}
\label{sp17} \left\{ \check {H}_f (x_f ;x_s ) - \check {\lambda}_i
(x_s) \right\} \Phi _i^{m\sigma} (x_f ;x_s ) = 0,
\end{equation}
in the interval $x_{f}\in \Omega_{x_f}(x_s)$, depending on the
conditionally slow variable $x_{s}\in\Omega_{x_s}$ as on a
parameter. These solutions obey the boundary conditions
\begin{equation}
\label{sp17a} \lim \limits_{x_f \to x_f^{t}(x_{s})}\!\! \left(\!\!
{N_f^{(m\sigma)}(x_{s}) g_{2f} (x_f)\frac{d\Phi _j^{m \sigma} (x_f
;x_s)}{dx_f } + D_f^{(m \sigma)}(x_{s}) \Phi _j^{m \sigma} (x_f;x_s
)}\!\!\right)\!\! =\!\! 0
\end{equation}
at the boundary points $\{x_f^{\min
}(x_{s}),x_f^{\max}(x_{s})\}=\partial\Omega_{x_f}(x_s)$, of the
interval $\Omega_{x_f}(x_s)$. In Eq. (\ref{sp17a}),
$N_f^{(m\sigma)}(x_{s})\equiv N_f^{(m\sigma)}$,
$D_f^{(m\sigma)}(x_{s})\equiv D_f^{(m\sigma)}$, unless specially
declared, are determined by the relations $N_f^{(m\sigma)} = 1$,
$D_f^{(m\sigma)} = 0$ at $m = 0$, $\sigma = + 1$ (or at $\sigma =0$,
i.e.,  without parity separation), $N_f^{(m\sigma)} = 0$,
$D_f^{(m\sigma)} = 1$ at $m = 0$, $\sigma = - 1$ or at $m \ne 0$.
The eigenfunctions satisfy the orthonormality condition with the
weighting function $g_{1f}(x_f)$ in the same interval $x_{f}\in
\Omega_{x_f}(x_s)$:
\begin{equation}
\label{sp19} \left\langle \Phi _i^{m\sigma} \vert \Phi _j^{m\sigma}
\right\rangle =
\int\nolimits_{x_f^{\min}(x_{s})}^{x_f^{\max}(x_{s})}\Phi
_i^{m\sigma} (x_f ;x_s ) \Phi _j^{m\sigma} (x_f ;x_s )
 g_{1f}(x_f)dx_f=\delta_{ij}.
\end{equation}
Here $\check \lambda_1 (x_s)< ... < \check \lambda_{j_{\max } }(x_s)
<...$ is the desired set of real eigenvalues. The corresponding set
of potential curves $ 2{E}_1 (x_s ) < ... < 2{E}_{j_{\max } } (x_s )
<...$ of Eqs. (\ref{sp09xf}) is determined by $2 {E}_{j} (x_s
)=g_{3s}^{-1} (x_s )\check \lambda_j (x_s)$. Note that for OSC and
PSC, the desired set of real eigenvalues $\check \lambda_j (x_s)$
depends on the combined parameter, $x_{s}\rightarrow
p^{2}=(d/2)^{2}2E$, i.e., the product of spectral $2E$ and geometrical
$(d/2)^{2}$ parameters of the problem (\ref{sp09}). The solutions of
the problem (\ref{sp17})--(\ref{sp19}) for Models A and B are
calculated in the analytical form \cite{CASC10}, while for Model C
this is done using the program  ODPEVP \cite{ODPEVP}.
 Substituting the expansion  (\ref{sp15}) into Eq. (\ref{sp01}),
we get a set of ODEs for the slow subsystem with respect to the
unknown vector functions $ {\mbox{\boldmath $\chi$}}^{(m\sigma
i)}(x_s ,E) \equiv {\mbox{\boldmath $\chi$}}^{(t)}(x_s ) = (\chi
_1^{(t)} (x_s ),...,\chi _{j_{\max } }^{(t)} (x_s ))^T$:
{\footnotesize\begin{eqnarray} &&
\label{sp23}
\!\!\!\!\!\!\biggl(\!
-\!\frac{1}{g_{1s}(x_{s})}\frac{d}{d{ x_{s}}}g_{2s}(x_{s})\frac{d}{d{ x_{s}}}
+\check {V}_{s}(x_s) +{V}_{ii}(x_s) - 2{E} \biggr)\chi_i^{(t)}({ x_{s}})
=\!-\!\sum_{j} {V}_{ij}(x_s)\chi_{j}^{(t)}({ x_{s}}).
\end{eqnarray}}
Here $  V_{ii}(x_s )= 2E_i (x_s )+H_{ii}(x_s )$, $  {V}_{ij}(x_s)$ are
defined by formula
{\footnotesize \begin{eqnarray}&&
 {V}_{ij}(x_s)= \frac{g_{2s}(x_s)}{g_{1s}(x_s)} H_{ij} ({ x_{s}})+
\frac{1}{g_{1s}(x_{s})}\frac{dg_{2s}(x_{s}) Q_{ij}({x_s})}{d{x_s}}
\!+\!\frac{g_{2s}(x_{s})}{g_{1s}(x_{s})} Q_{ij}({ x_{s}})
\frac{d}{d x_{s}},\nonumber
\\&&
H_{ij}(x_{s})=H_{ji}( x_{s})=
\int\nolimits_{x_f^{\min}(x_{s})}^{x_f^{\max}(x_{s})}g_{1f}(x_{f})
\frac{\partial\Phi_{i}(x_{f}; x_{s})}{\partial x_{s}}
\frac{\partial\Phi_{j}(x_{f}; x_{s})}{\partial x_{s}}dx_{f},
\label{sp23a}  \\&&
Q_{ij}(x_{s})=-Q_{ji}( x_{s})=
-\int\nolimits_{x_f^{\min}(x_{s})}^{x_f^{\max}(x_{s})}g_{1f}(x_{f})
\Phi_{i}(x_{f}; x_{s})
\frac{\partial\Phi_{j}(x_{f}; x_{s})}{\partial x_{s}}dx_{f},
 \nonumber
\end{eqnarray}}
and calculated analytically for Model B
 and by means of the program
ODPEVP \cite{ODPEVP} for Model C,
while the solutions of the BVPs for Eqs. (\ref{sp23})
with the boundary and  orthonormalization conditions  of the type (\ref{sp17a}), (\ref{sp19})  with
$x_{f}\rightarrow x_{s}$ were calculated by means of the program
KANTBP \cite{kantbp}. Note, that for Model A in SC or CC and Model B in OSC or PSC,
the variables $x_f$ and $x_s$ are separated so that the matrix elements
$\check V_{ij} (x_s) 0$ are put into the r.h.s. of Eq. (\ref{sp23}),
and $V_s(x_s)$ are substituted from Table \ref{ktpy}.
For the interesting lower part of the spectrum of Models A and B $2
E:    2 E_1 < 2 E_2 < \ldots < 2 E_t $, or of Model C $2 E:    2 E_1
< 2 E_2 < \ldots < 2 E_t < 2 U_{0}$, the number $j_{\max}$ of the
equations solved
should be at least not less than the number of the
energy levels of the problem (\ref{sp23}) at $a=c=r_{0}$.
To ensure the prescribed
accuracy of calculation of the lower part of the spectrum discussed below with
eight significant digits we used $j_{\max}= 16$ basis functions in the expansion (8) and
the discrete approximation of the desired solution by Lagrange finite elements
of the fourth order with respect to the grid pitch
$\Omega^p_{h^s(x_s)}=[x_{s;\min}, x_{s;k} = x_{s;k-1}+h_s,x_{s;\max}]$.
The details of the corresponding
computational scheme are given in~\cite{CASC10}.

\section{Spectral Characteristics of Spheroidal \\ and Dumbbell QDs}\label{3}
\subsection{Model A of OSQD \& PSQD }\label{31}
In the exactly solvable model A the variables are separable in
spherical coordinates, and under the variation of the aspect ratio
parameters $\zeta_{ca}=c/a$ and $\zeta_{ac}=\zeta_{ca}^{-1}=a/c$
for the oblate and prolate spheroids, determining the transverse $
\omega_{\rho}=\sqrt\zeta_{1} \omega$ and longitudinal $
\omega_{z}=\sqrt\zeta_{3} \omega$ frequencies of the circular and
linear harmonic oscillators. The spectrum is given by the sum of
energies $2 E_{n_{\rho}m}=2\omega_{\rho}(2n_{\rho}+|m|+1),
n_{\rho}=0,1,\ldots, m=0,\pm1,\ldots$ (with the eigenvalues being
degenerate with respect to $\lambda_{\rho}=2n_{\rho}+|m|$ that
number in ascending order the energy values of the states
\cite{MPST06,KKWP02} that is conventionally used in practice, see, for example,
\cite{Rassey58,Poenaru08})  and $2 E_{n_{z}}=2\omega_{z}(n_{z}+1/2),
n_{z}=0,1,\ldots,$
 at $ \omega= \omega_{r_{0}}=\pi^2/(3   r_{0}^2)$,
$\sqrt\zeta_{1}=  r_{0}^{2}/(a^2)$, and $\sqrt\zeta_{3}=
r_{0}^{2}/(c^2)$. At $a=c=r_{0}$ the independent variables are
separable in the boundary problem for Eq.  (\ref{sp01}) in the spherical
coordinates too, i.e., we have the energy spectrum of a spherical
oscillator $2 E_{n_{r}lm}^{osc}=2\omega_{r_{0}}(2n_{r}+l+3/2)$,
$n_{r}=0,1,\ldots$, $l=0,1,\ldots$, $m=0,\pm 1,\ldots,\pm l$ with
the eigenvalues being degenerate with respect not only  to $m$, but
also to $\lambda_r=2n_{r}+l$ that number in ascending order the energy
values of states, separated in parity
$\hat{\sigma}=(-1)^{\lambda}=(-1)^l=(-1)^{m}\sigma$,
$\sigma=(-1)^{l-m}=\pm1$. The energy spectrum of the spherical
oscillator $2 E_{n_{r}lm}^{osc} $
coincides at $a=c$ with
\begin{eqnarray} 2 E(a,c)=2( E_{n_{z o}}+ E_{n_{\rho o}m}) \quad and
\quad 2 E(c,a)=2( E_{n_{\rho p},m}+ E_{n_{z p}}), \label{osc81}
\end{eqnarray} which,
respectively, defines the one-to-one correspondence between the sets
of the quantum numbers
$n_{z o}=l-|m|$, $n_{\rho o}=n_{r}$, $m=m$ for OSQD and SQD and $n_{\rho p}=n_{r}$, $m=m$,
$n_{z p}=l-|m|$
for PSQD and SQD, that characterize the fast and slow subsystems at continuous variation of
the parameters $\zeta_{ca}=c/a$ and $\zeta_{ac}=a/c$.
At decreasing the parameter  $\zeta_{ca}$ or $\zeta_{ac}$
the degeneracy of the spectrum with respect to the quantum numbers  $n$, $l$, $m$ is removed.

\begin{figure}[t]
\includegraphics[width=0.49\textwidth,height=0.4\textwidth]{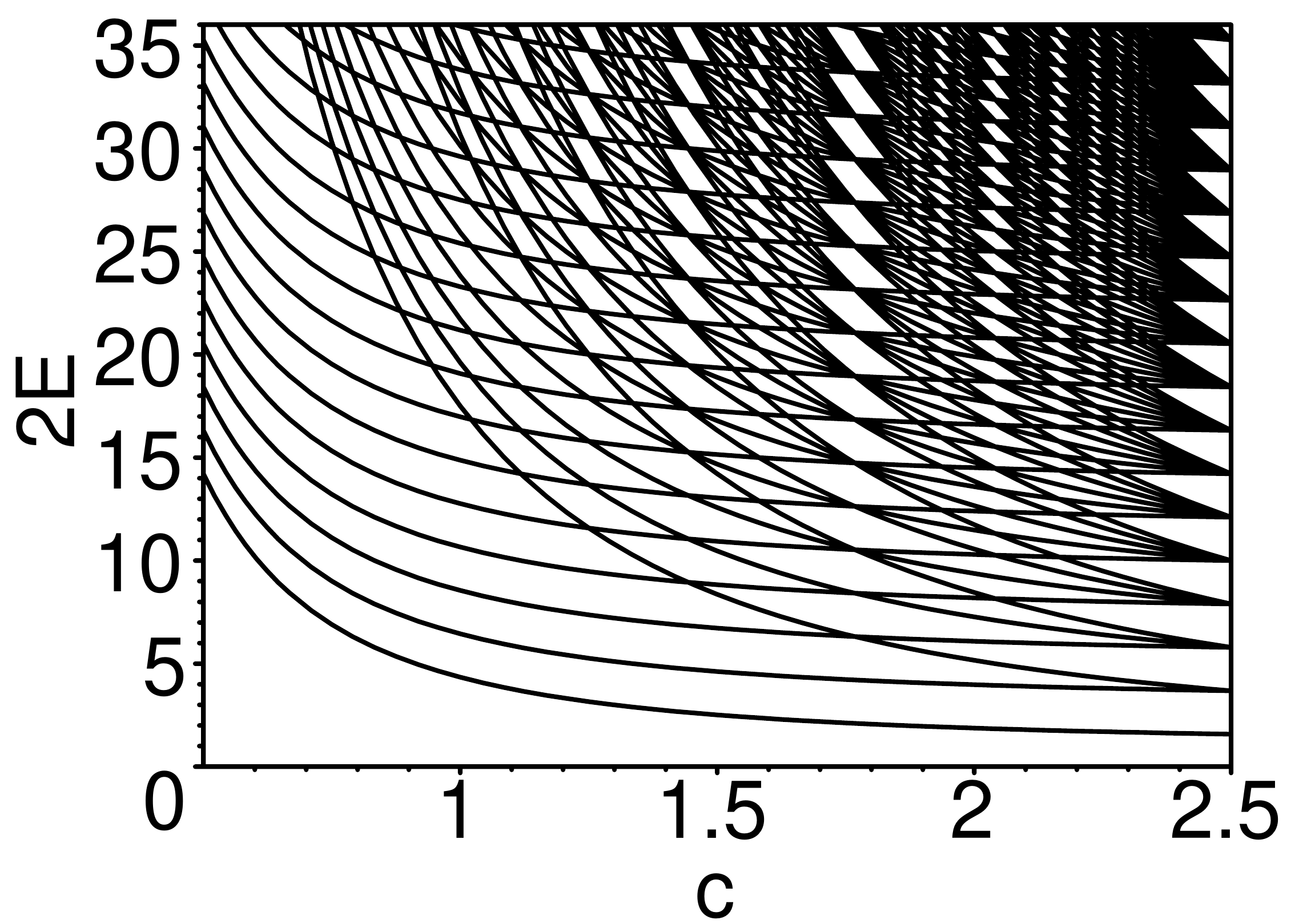} \hfill
\includegraphics[width=0.49\textwidth,height=0.4\textwidth]{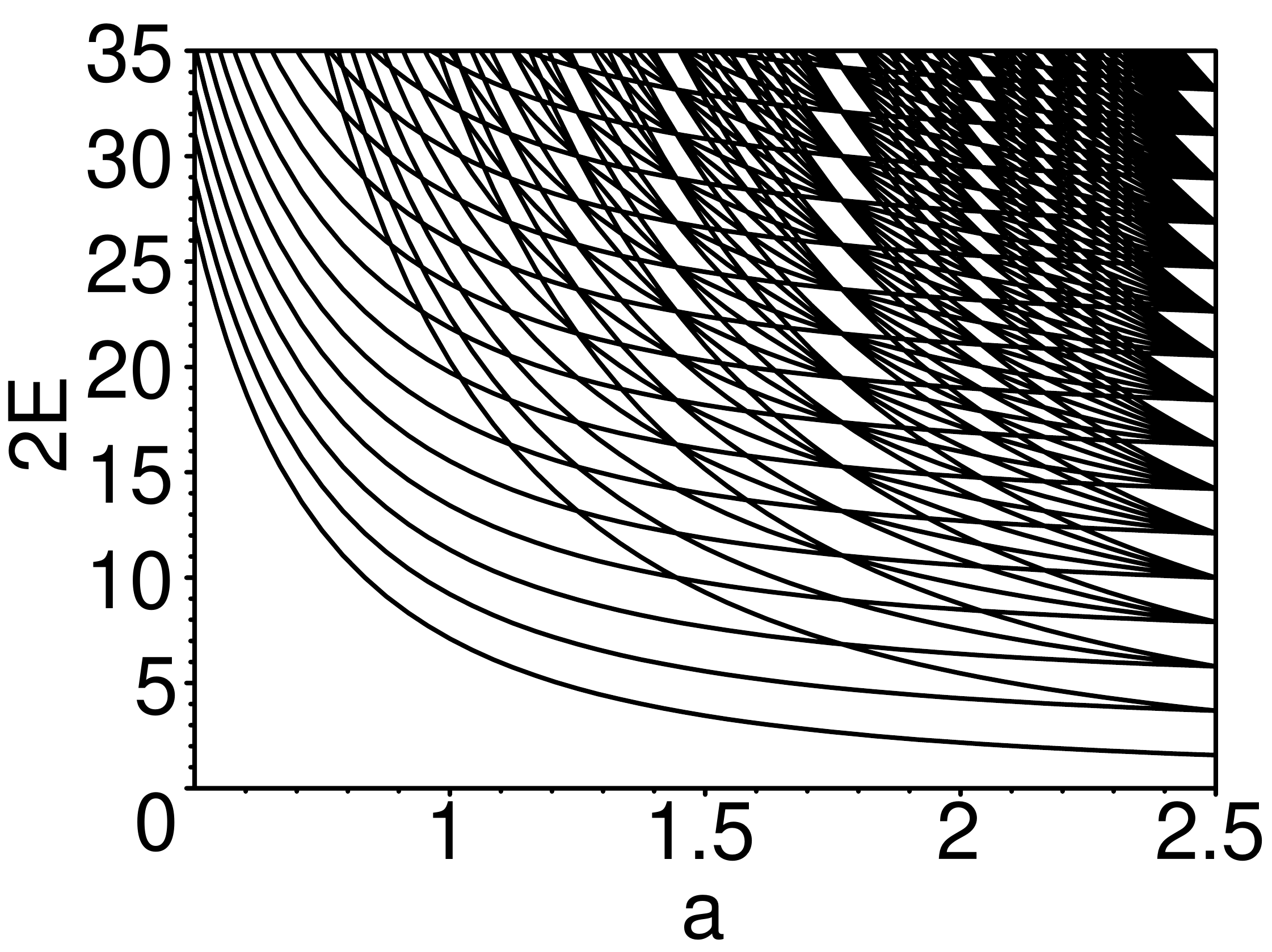}\\
 \parbox{0.14\textwidth}{\phantom{a}}
 \parbox{0.49\textwidth}{a)}
 \parbox{0.25\textwidth}{b)}
\caption{
Energies $2 E=\tilde E/E_R$ of the even $\sigma=+1$ lower
states of Model A OSQD at $a=2.5$ (a) and PSQD at $c=2.5$ (b) versus
$c$ or $a$.
{
The exact intersections of the energy levels take place at rational
ratios $R=\omega_{\rho}/\omega_{z}=(c/a)^2\in {\cal Q}$ (a)
and $R=\omega_{z}/\omega_{\rho}=(a/c)^2\in {\cal Q}$ (b)
of the frequencies of transverse and
longitudinal oscillators with $R=1,4/5,3/4,2/3,3/5,1/2,2/5,1/3,1/4,1/5,...$}
} \label{occs}
\end{figure}

{Fig. \ref{occs} illustrates the lower part of the equidistant
energy spectrum $\tilde E/\tilde E_{R}=2  E (a,c)$
and $\tilde E/\tilde E_{R}=2   E (c,a)$ for even states $\sigma=+1$
of the model of OSQD and PSQD with parabolic confining potentials
(\ref{sp02}), at $m=0$, i.e., of an oblate and prolate spheroid,
depending on the minor  $c$ or $a$ and the major $a$ or $c$
semiaxes, respectively. } At fixed  values of the parity $\sigma$
and the magnetic quantum number $m$ when the ratio of the
frequencies $\omega_{\rho}$ and $\omega_{z}$ of the longitudinal and
transverse oscillators is a rational number,
$\omega_{\rho}/\omega_{z}\in {\cal Q}$, as illustrated, e.g.,  in
Fig. \ref{occs}, the exact crossings of the same-parity terms occur,
after which above each energy level of OSQD (or PSQD), labelled with
the quantum number $n_{z o}$ (or $n_{\rho p}$) of the fast
subsystem, an equidistant spectrum appears with the energy levels
labelled with the quantum number $n_{\rho o}$ (or $ n_{z p}$) of the
slow subsystem. Note, that when the parameters tend to zero, the
longitudinal energy of OSQD and the transverse energy of PSQD tend
to infinity. However, since the variables are separable and the
energy can be presented as a sum, the finite energies for a disc
$E_{n_{\rho o}m}$ or a wire $E_{n_z p}$ result from the subtraction
of the longitudinal  $E_{n_{z o}}$ or transverse $E_{n_{\rho p}m}$
energy, respectively.

\subsection{Models B and C for Oblate Spheroidal QD.}
At a fixed coordinate  $x_{s}$ of the slow subsystem, the motion of
the particle in the fast degree of freedom $x_{f}$ is localized
within the potential well having the effective width
\begin{equation}
\label{eq70}
  L\left( x_s \right) = 2c\sqrt {1 - {x_s^2}/{a^2}} ,
\end{equation}
where $L=\tilde L/a_B^*$. The parametric BVP for Eq. (\ref{sp09xf})
at fixed values of the coordinate  $x_{s}$, $x_{s}\in (0,a)$, is
solved in the interval $x_f\in(- L\left( x_s \right)/2, L\left( x_s
\right)/2)$ for Model C using the program ODPEVP, and for Model B
the eigenvalues $\tilde E_{n_{o}}\left( x_s \right)/\tilde E_R\equiv
2 E_{i}\left( x_s \right)$, $n_{o}=i = 1,2,...$, and the
corresponding parametric eigenfunctions $\Phi^\sigma_{i} \left(
{x_f;x_s}\right)$, are expressed in the analytical form:
\begin{equation}
\label{eq71}
\!\!\!\!\!\!\!\!2 E_{i}\left( x_s \right)\!=\!\frac{\pi ^2n_{o}^2}{
L^2\left( x_s \right)},\quad \Phi_{i}^{\sigma} \left(
{x_f;x_s}\right)\!=\!\sqrt{\frac{2}{ L\left( x_s \right)}}
\sin\left(\frac{\pi n_{o}}{2} \left(\frac{x_f}{ L\left( x_s
\right)/2}-1\right)\right),
\end{equation}
where the even solutions $\sigma=+1$  are labelled with odd $n_{o}=n_{zo}+1=2i-1,$
and the odd ones $\sigma=-1$
with even $n_{o}=n_{zo}+1=2i$, $i=1,2,3,...$ .
The effective potentials  (\ref{sp23a}) in Eq. (\ref{sp23})
for the slow subsystem are expressed analytically
in terms of the integrals over the fast variable  $x_f$ of the basis functions (\ref{eq71})
and their derivatives with respect to the parameter $x_s$
{including the states of both parities $\sigma=\pm1$}:
\begin{eqnarray}
&& 2  E_{i}(x_{s})=
\frac{a^2\pi^2n_{o}^2}{4c^2(a^2-x_s^2)},\quad
H_{ii}(x_{s})=\frac{3+\pi^2n_{o}^2}{12}\frac{x_s^2}{(a^2-x_s^2)^2},\label{sp23s}\\
&&H_{ij}(x_{s})=\frac{2n_{o}n_{o}'(n_{o}^2+n_{o}'{}^2)(1+(-1)^{n_{o}+n_{o}'})}{(n_{o}^2-n_{o}'{}^2)^2}\frac{x_s^2}{(a^2-x_s^2)^2},
\nonumber\\
&&Q_{ij}(x_{s})=\frac{n_{o}n_{o}'(1+(-1)^{n_{o}+n_{o}'})}{(n_{o}^2-n_{o}'{}^2)^2}\frac{x_s}{a^2-x_s^2},\quad n_{o}'\neq n_{o}.\nonumber
\end{eqnarray}

\begin{figure}[t]
\includegraphics[width=0.44\textwidth,height=0.4\textwidth]{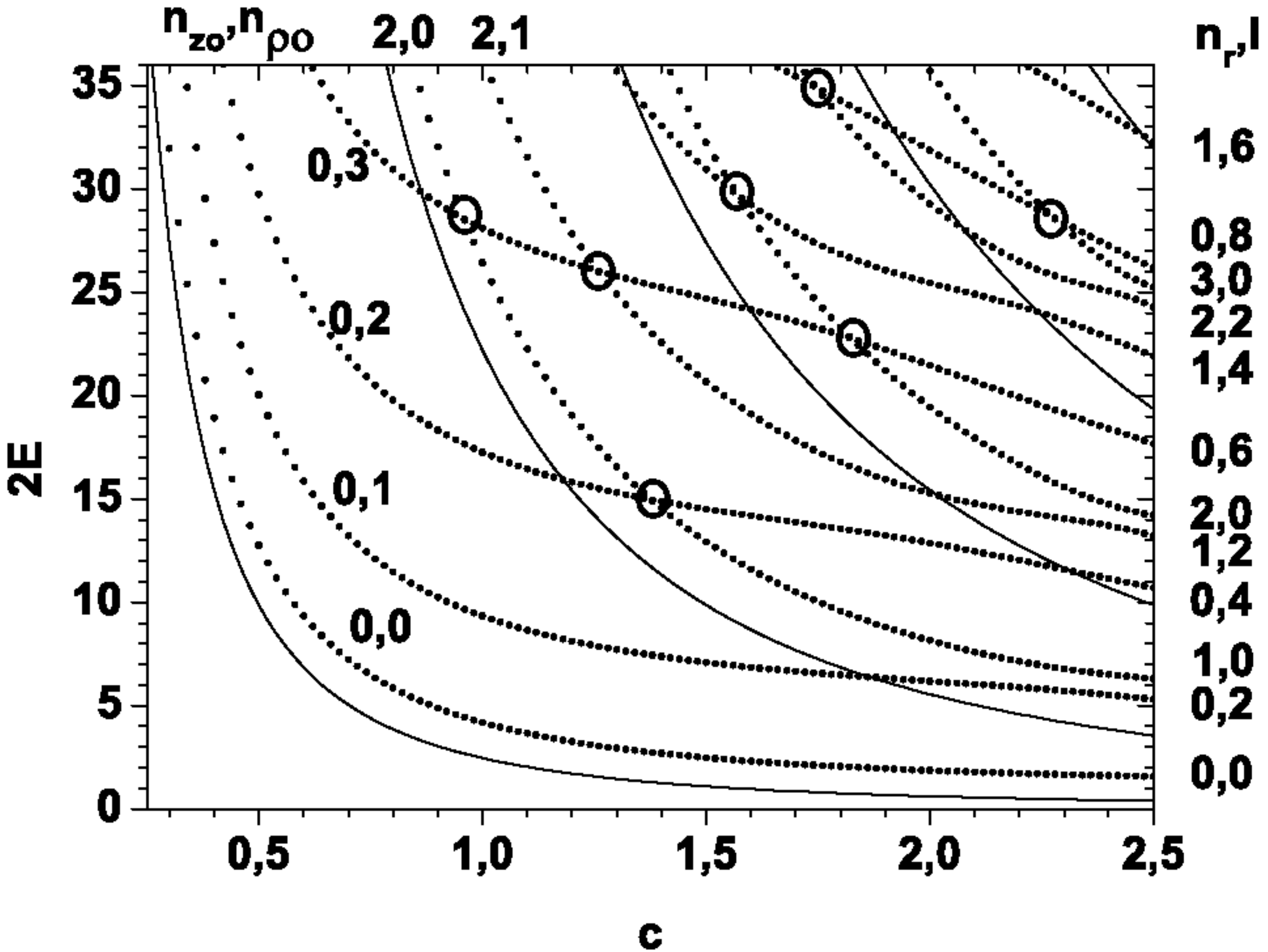}a) \hfill
\includegraphics[width=0.44\textwidth,height=0.4\textwidth]{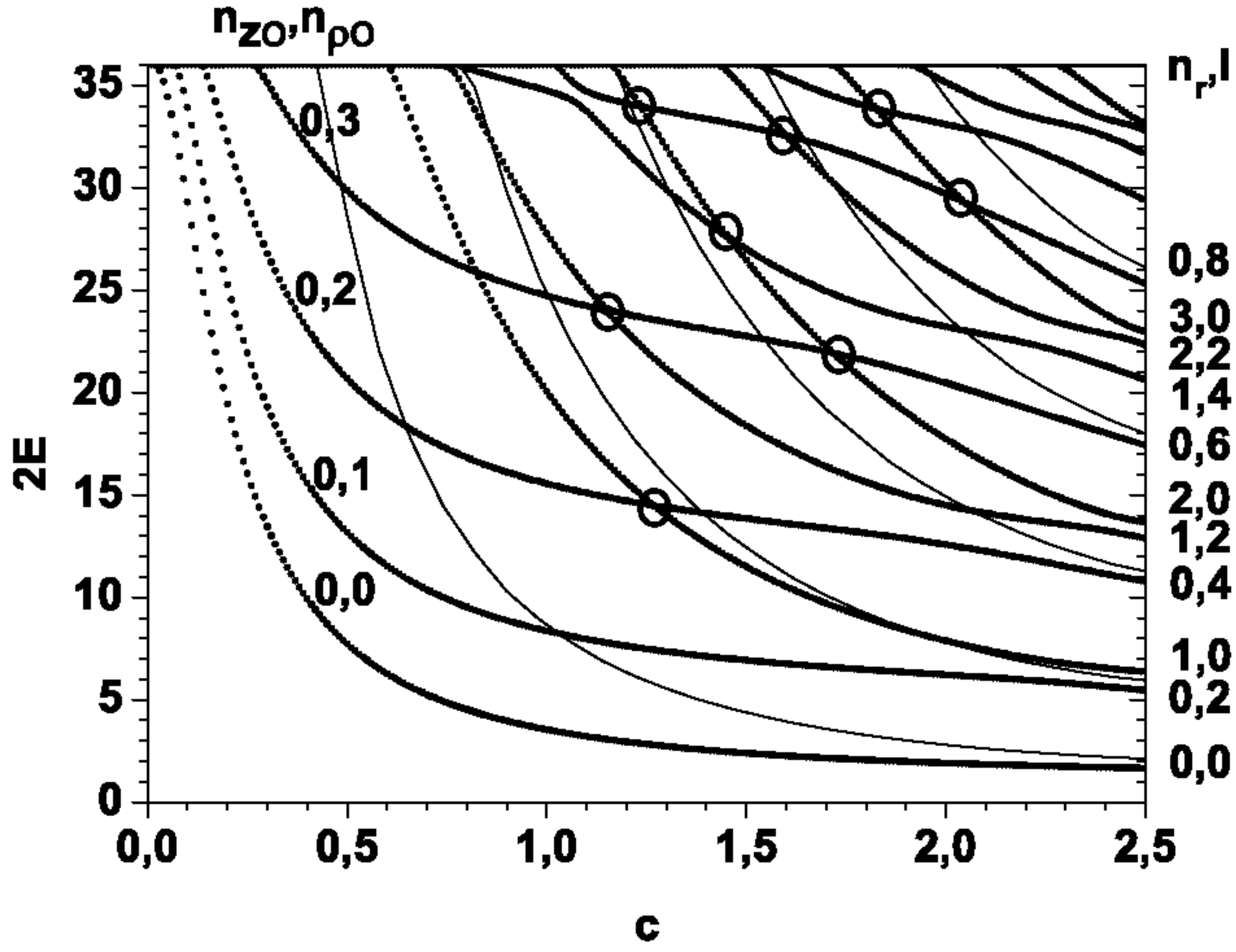}b)\\
\caption{The energies $2E=\tilde E/E_R$ of even $\sigma=+1$ lower states
for OSQD versus the minor $c$, $\zeta_{ca}=c/a\in(1/5,1)$ being the
spheroid aspect ratio: a) well with impermeable walls, b)
diffusion potential with $2U_0=36$, $s=0.1$,  the major semiaxis
$a=2.5$ and $m=0$. Tine lines are minimal values
$2E_{i}^{min}\equiv 2E_{i}(x_{s}=0)$ of potential curves.} \label{enekts}
\end{figure}

\begin{figure}[t]
\includegraphics[width=0.9\textwidth]{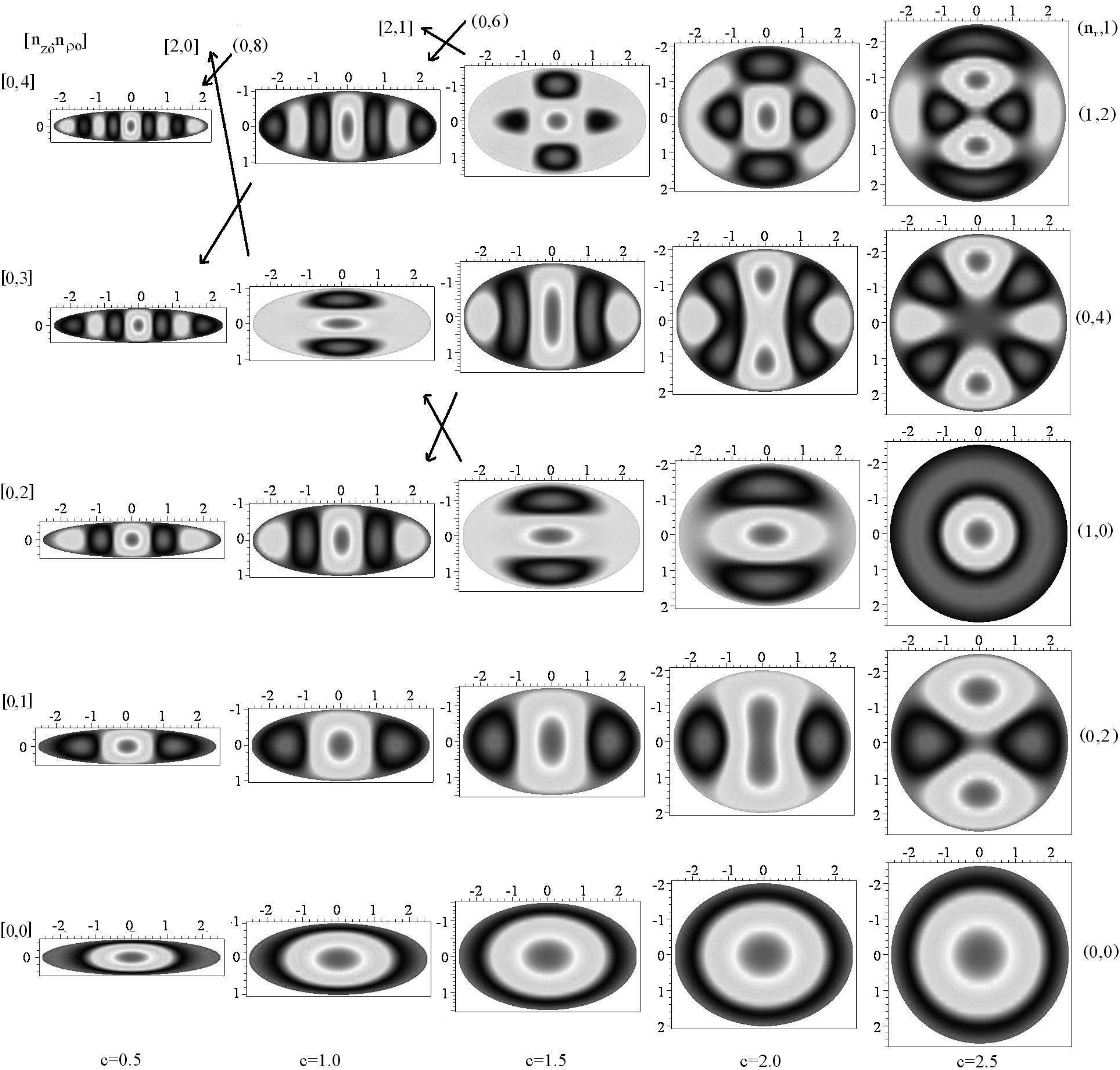}
\caption{Contour lines of the  first five even-parity wave
functions $\sigma=+1$ in the $xz$ plane of Model B of OSQD for
the major semiaxis $a=2.5$ and different values of the minor
semiaxis $c$ ($\zeta_{ca}=c/a\in(1/5,1)$) } \label{cwe}
\end{figure}

For Model B at  $c=a=r_{0}$ the OSQD turns into  SQD with known analytically expressed
energy levels $E_{t}\equiv E_{nlm}^{sp}$ and the corresponding eigenfunctions
\begin{equation}\label{exact}
2 E_{nlm}^{sp}\!=\!\frac{\alpha_{n_r+1,l+1/2}^2}{r_{0}^2},~
\Phi_{nlm}^{sp}(r,\theta,\varphi)
\!=\!\frac{\sqrt{2}J_{l+1/2}(\sqrt{2 E_{nlm}^{sp}}r)}
{r_{0}\sqrt{r}|J_{l+3/2}(\alpha_{n_r+1,l+1/2})|}
Y_{lm}(\theta,\varphi),
\end{equation}
where $\alpha_{n_r+1,l+1/2}$ are zeros of the Bessel function of
semi-integer index $l+1/2$, numbered in ascending order
$0<\alpha_{11}<\alpha_{12}< ...<\alpha_{iv}<...$ by the integer $i,v=1,2,3,...$.
Otherwise one can use
equivalent pairs $iv\leftrightarrow\{n_r,l\}$ with
 $n_r=0,1,2,...$ numbering the zeros of the Bessel function and  $l=0,1,2,...$,
being  the orbital quantum number that determines the parity of
states $\hat{\sigma}=(-1)^l=(-1)^{m}\sigma$,
$\sigma=(-1)^{l-m}=\pm1$. At fixed  $l$,  the energy levels
$\tilde E_{nlm}/\tilde E_{R}=2 E_{t}$ degenerate with respect to the
magnetic quantum number $m$, are labelled with the quantum number
$n=n_r+1=i=1,2,3,...$ , in contrast to the spectrum of a spherical
oscillator, degenerate with respect to the quantum number
$\lambda=2n_{r}+l$. Figs. \ref{enekts}, \ref{cwe} show the lower
part of the non-equidistant spectrum $\tilde
E(\zeta_{ca})/\tilde E_{R}=2  E_t$ and the eigenfunctions
$\Psi^{m\sigma}_t$ from Eq. (\ref{sp15}) for even states OSQD Models
B and C at $m=0$. There is a one-to-one correspondence rule
$n_{o}=n_{zo}+1=2n-(1+\sigma)/2, n=1,2,3,...$, between the sets of
spherical quantum numbers  $(n,l,m,\hat\sigma)$ of SQD with radius
$r_0=a=c$ and spheroidal ones  $\{n_\xi=n_r,n_\eta=l-|m|,m,\sigma\}$
of OSQD with the major $a$ and the minor $c$ semiaxes, and the
adiabatic set of cylindrical quantum numbers $[n_{zo},n_{\rho
o},m,\sigma]$ at continuous variation of the parameter
$\zeta_{ca}=c/a$.
 The presence of crossing points of the energy levels of similar parity
under the symmetry change from spherical $\zeta_{ca}=1$ to axial,
i.e., under the variation of the parameter $0<\zeta_{ca}<1$, in
the BVP with two variables at fixed  $m$ for Model B is caused by
the possibility of variable separation for Eq. (\ref{la1949o}) in the OSC \cite{stigun},
i.e., the r.h.s. of Eq. (\ref{sp23} ) equals zero,
and by the existence of the integral of motion (\ref {la1949}).
The transformation of the eigenfunctions occurring in the course of a
transition through the crossing points (marked by circles) in Fig.
\ref{enekts}, is shown in Fig. \ref{cwe} for model B (marked by arrows) and
similar for model C. From the comparison of
these Figures one can see that if the eigenfunctions are ordered
in accordance with the increasing eigenvalues of the BVPs, then for both
Models B and C, the number of nodes~\cite{CurantGilbert} is
invariant under the variation of the parameter $c$ from $c=a=2.5$ to
$c=0.5$ of the potentials (\ref{sp03}) and (\ref{sp03VS}). For Model
B, such a behavior follows from  the fact of separation of
variables of the BVP with the potential (\ref{sp03}) in the OSC,
while for Model C further investigation is
needed, because the coordinate system, in which the variables of the
BVP with the potential (\ref{sp03VS}) are separable,
is unknown.
So, at small values of the deformation parameter ($\zeta_{ca}$ for OSQD or
$\zeta_{ac}$ for PSQD) there are nodes only along the corresponding
major semiaxis.
For Model C at each value of the
parameter $a$ there is a finite number of discrete energy levels
limited by the value $2U_0$ of the well walls height. As shown in
Fig. \ref{enekts}b, the number of levels of OSQD, equal to that of
SQD at $a=c=r_0$, is reduced with the decrease of the parameter
$c$ (or $\zeta_{ca}$), in contrast to Models A and B that have
countable spectra, and avoided crossings appear just below the
threshold.

\subsection{Models B and C for Prolate Spheroidal QD.}
In contrast to OSQD, for PSQD at fixed coordinate  $x_s$
of the slow subsystem the motion of the particle
in the fast degree of freedom $x_{f}$
is confined to a 2D potential
well with the effective variable radius
\begin{eqnarray}
\label{eq99}
  \rho_{0} \left( x_{s};a,c \right) = a\sqrt {1 - {x_s^2} / {c^2}} ,
\end{eqnarray}
where $ \rho_{0}\left( x_{s} \right) = \tilde \rho_{0}\left( x_{s}
\right)/ {a_B^{*} } $. The parametric BVP for  Eq. (\ref{sp09xf}) at
fixed values of the coordinate $x_{s}$ from the interval $x_{s}\in
(-c,c)$ is solved in the interval $x_f\in(0, \rho_{0}\left( x_{s}
\right))$ for Model C using the program ODPEVP, while for Model B
the eigenvalues
$\tilde E_{n_{\rho p}+1}\left( x_s \right)/\tilde E_R\equiv
2E_{i}\left( x_s \right)$, $n_{\rho p}+1=i = 1,2,...$, and the
corresponding parametric basis functions $\Phi^{m\sigma=0}_{i}
\left( {x_f;x_s}\right)\equiv\Phi^{m}_{i} \left( {x_f;x_s}\right)$
without parity separation are expressed in the analytical form:
\begin{equation}
\label{eq100}
2 E _i \left( x_s \right) = \frac{\alpha _{n_{\rho p} + 1,|m|}^2}
 {  \rho_{0}^2\left( x_s \right)},\quad
\Phi^{m}_{n_{\rho p}}(x_{s})=\frac{\sqrt{2}}{  \rho_0\left( x_s
\right)} \frac{ J_{|m|}(\sqrt{2  E _{n_{\rho p}+1,|m|} \left( x_s
\right)}x_{f}) }{|J_{|m|+1}(\alpha_{n_{\rho p}+1,|m|})|},
\end{equation}
where $\alpha _{n_{\rho p}+ 1,|m|}=\bar J^{n_{\rho p}+ 1}_{|m|}$
are positive zeros of the Bessel function of the first kind
$J_{|m|}(x_f)$, labeled in the ascending order with the quantum
number $n_{\rho p}+1 =i= 1,2,...$.

The effective potentials (\ref{sp23a}) in Eq.(\ref{sp23}) for the
slow subsystem are calculated numerically in  quadratures via the
integrals over the fast variable  $x_f$ of the basis
functions(\ref{eq100}) and their derivatives with respect to the
parameter  $x_s$, and at $m=0$ may be presented in the analytical form:
\begin{eqnarray}
&&
2  E_i \left( x_s \right) = \frac{(\bar
J_0^i)^2}{\rho_{0}^2\left( x_s \right)}, \quad
H_{ii}(x_{s})=\left(\frac{\rho_{0}'\left( x_s \right)}{\rho_{0}\left( x_s \right)}\right)^2\frac{ (1+\bar J_0^i)}{3 },\label{sp23v}\\
&&H_{ij}(x_{s})=
2\left(\frac{\rho_{0}'\left( x_s \right)}{\rho_{0}\left( x_s \right)}\right)^2
\left(\bar J_0^i\bar J_0^j\int_0^1 \frac{J_1(\bar J_0^ix)}{J_1(\bar J_0^i)}
 \frac{J_1(\bar J_0^jx)}{J_1(\bar J_0^j)}x^3dx\right.
 \nonumber \\&&\left.\qquad\qquad
-
\bar J_0^i\int_0^1 \frac{J_1(\bar J_0^ix)}{J_1(\bar J_0^i)}
 \frac{J_0(\bar J_0^jx)}{J_1(\bar J_0^j)}x^2dx
-\bar J_0^j\int_0^1 \frac{J_0(\bar J_0^ix)}{J_1(\bar J_0^i)}
 \frac{J_1(\bar J_0^jx)}{J_1(\bar J_0^j)}x^2dx
\right) ,\nonumber\\
&&Q_{ij}(x_{s})=-2\frac{\rho_{0}'\left( x_s \right)}{\rho_{0}\left( x_s \right)}\bar J_0^j\int_0^1 \frac{J_0(\bar J_0^ix)}
{J_1(\bar J_0^i)}\frac{J_1(\bar J_0^jx)}{J_1(\bar J_0^j)}x^2dx,\quad j\neq i.
\nonumber
\end{eqnarray}

\begin{figure}[t]
\includegraphics[width=0.44\textwidth,height=0.4\textwidth]{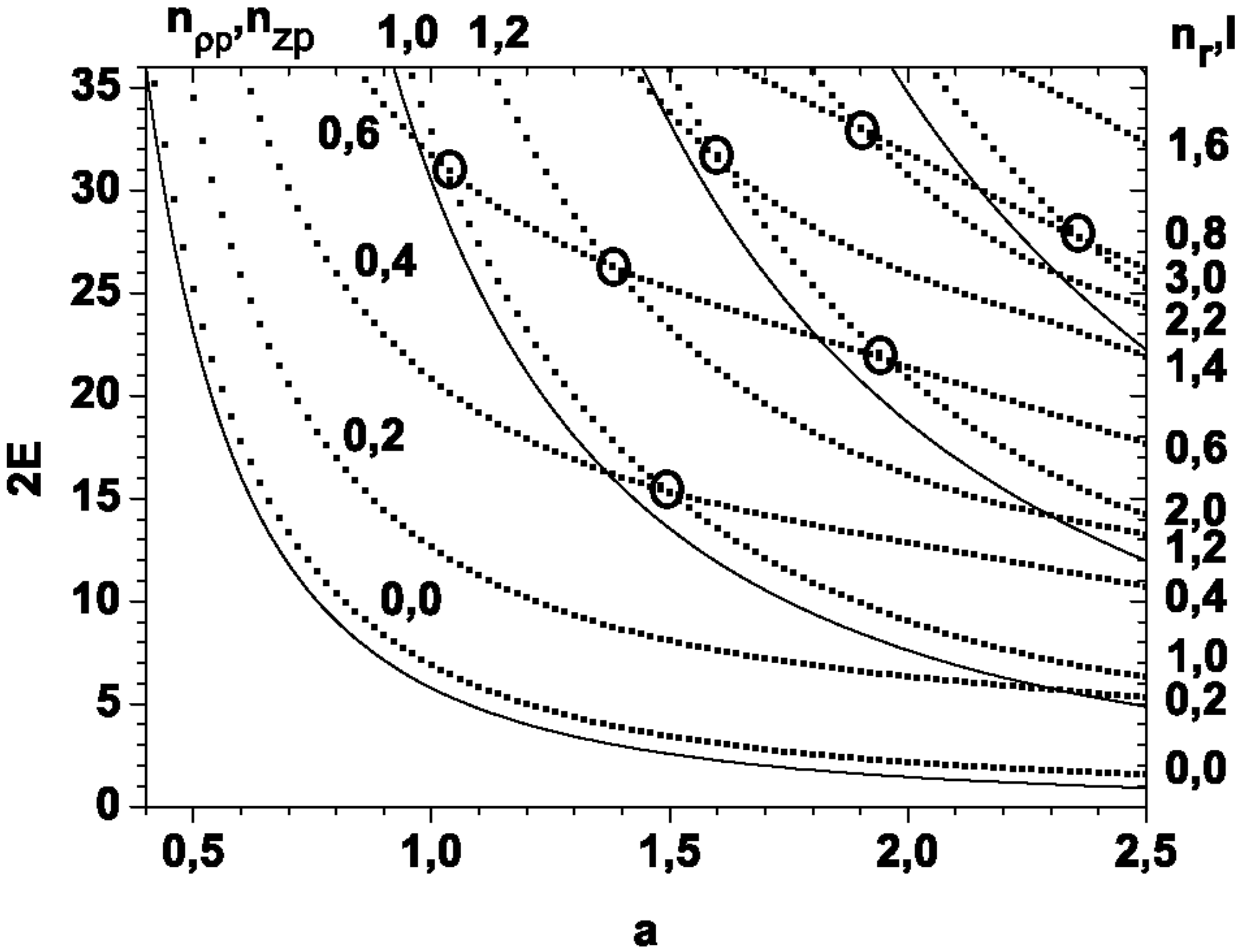} a) \hfill
\includegraphics[width=0.44\textwidth,height=0.4\textwidth]{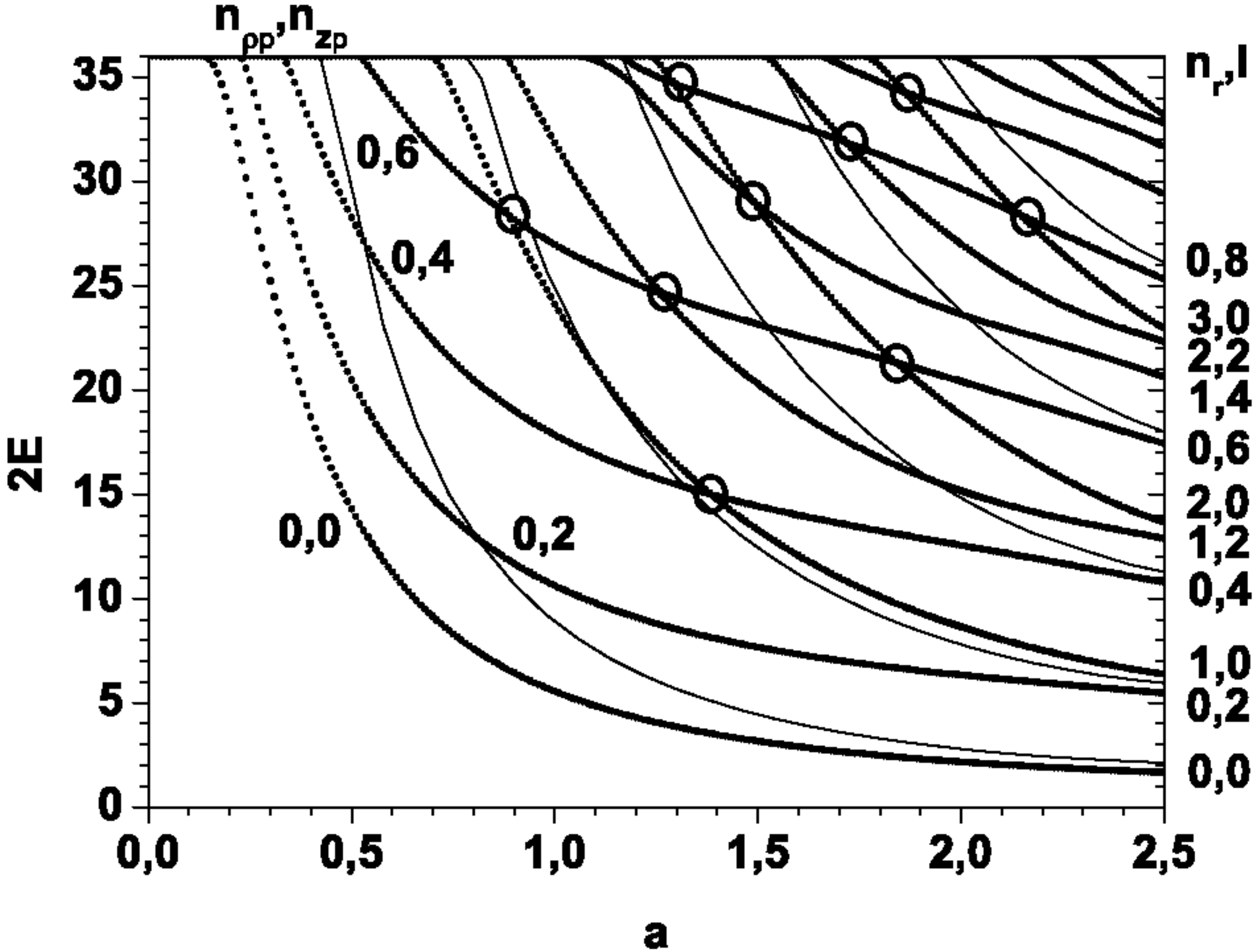}b)\\
\caption{The energies $2E=\tilde E/E_R$  of even $\sigma=+1$
lowest states for PSQD depending on the minor semiaxis  $a$
$(\zeta_{ac}=a/c\in(1/5,1)$ is the spheroid aspect ratio): a) well
with impermeable walls, b) diffusion potential,    $2U_0=36$,
$s=0.1$, for the major semiaxis $c=2.5$ and $m=0$. Tine lines are
minimal values $2E_{i}^{min}\equiv 2E_{i}(x_{s}=0)$ of potential
curves} \label{enektv}
\end{figure}

\begin{figure}[t]
\includegraphics[width=0.9\textwidth]{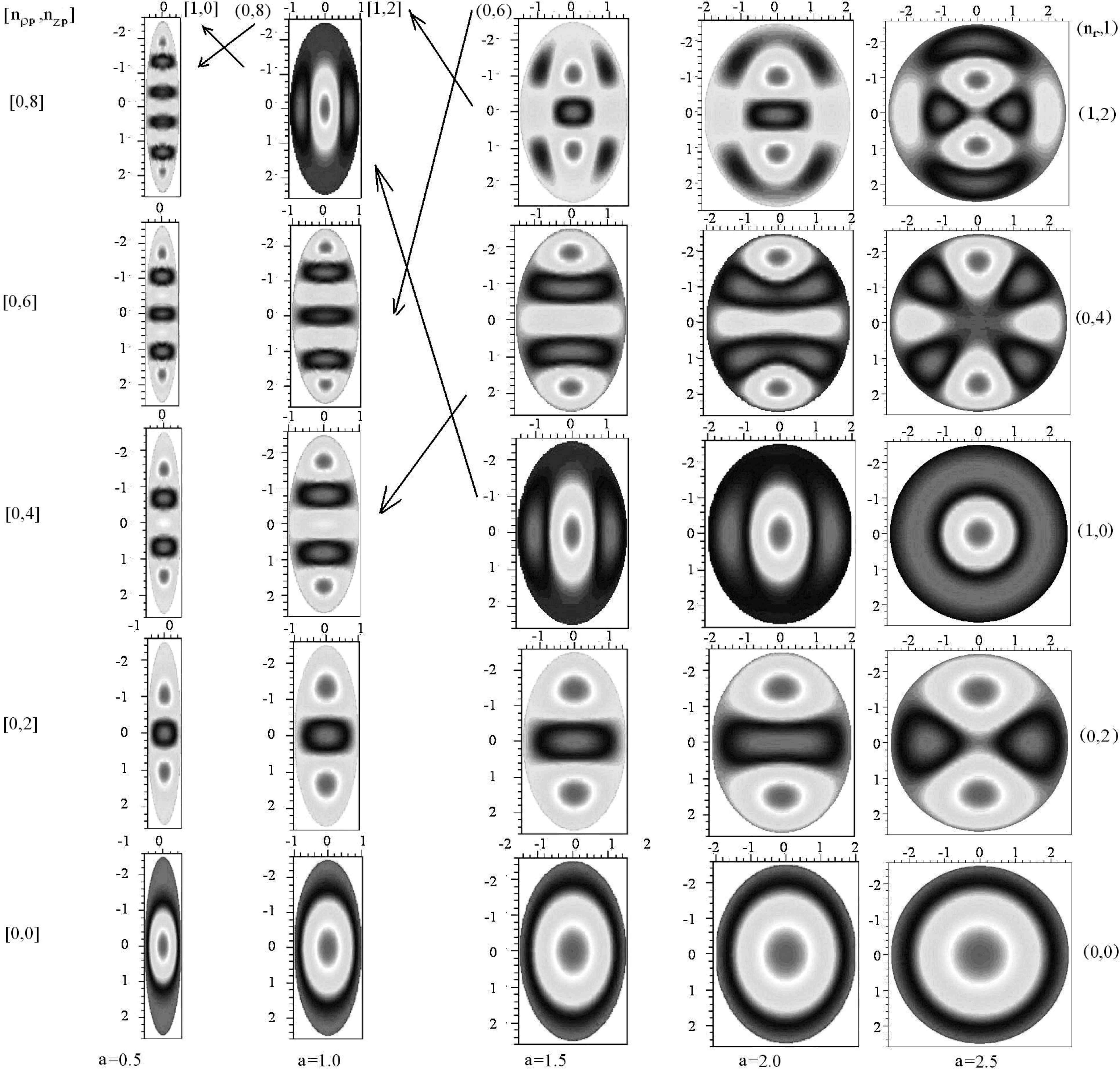}
\caption{Contour lines of the  first five even-parity wave
functions $\sigma=+1$ in the $xz$ plane of Model B of PSQD for
the major semiaxis $c=2.5$ and different values of the minor
semiaxis $a$ ($\zeta_{ac}=a/c\in(1/5,1)$)
} \label{awe}
\end{figure}

Figures \ref{enektv}, \ref{awe}
illustrate the lower part of the
non-equidistant spectrum $\tilde E(\zeta_{ac})/\tilde E_{R}$ = $2
E_t$ and the eigenfunctions $\Psi^{m\sigma}_t$ from Eq. (\ref{sp15})
of even states of PSQD Models B and C.

A one-to-one
correspondence rule  $n_{\rho p}+1=n_p=i=n=n_r+1$, $i=1,2,...$ and
$n_{z p}=l-|m|$ holds between the  quantum numbers
$(n,l,m,\hat\sigma)$ of SQD with the radius  $r_0=a=c$, the
spheroidal quantum numbers $\{n_\xi=n_r,n_\eta=l-|m|,m,\sigma\}$ of PSQD with
the major $c$ and the minor $a$ semiaxes, and the adiabatic set of
quantum numbers $[n_p=n_{\rho p}+1,n_{z p},m,\sigma]$ under the
continuous variation of the parameter $\zeta_{ac}=a/c$.
 The presence of crossing points of similar-parity energy levels in
Fig.   \ref{enektv} under the change of symmetry from spherical
$\zeta_{ac}=1$ to axial, i.e., under the variation of the
parameter $0<\zeta_{ac}<1$, in the BVP with two variables at fixed
$m$ for Model B is caused by the possibility of variable
separation for Eq. (\ref{la1948o}) in the PSC ~\cite{stigun},
i.e., r.h.s. of Eq. (\ref{sp23}) equals zero,
and by the existence of the additional integral of motion (\ref{la1948}).
For Model C, at each value of the
parameter $c$ there is also only a finite number of discrete
energy levels limited by the value $2 U_0$ of the well walls
height. As shown in Fig. \ref{enektv}b, the number of energy
levels of PSQD, equal to that of SQD at $a=c= r_0$, which is
determined by the product of mass $\mu_e$ of the particle, the
well depth $\tilde U_0$, and the square of the radius $ \tilde
r_0$, is reduced with the decrease of the parameter $\tilde a$ (or
$\zeta_{ac}$) because of the promotion of the potential curve
(lower bound) into the continuous spectrum, in contrast to Models
A and B having countable spectra. Note, that the spectrum of Model
C for PSQD or OSQD should approach that of Model B with the growth
of the walls height $U_0$ of the spheroidal well. However, at
critical values of the ellipsoid aspect ratio it is shown that in
the effective mass approximation, both the terms (lower bound) and
the discrete energy eigenvalues in models of the B type are shifter towards
the continuum. Therefore, when approaching the critical aspect
ratio values, it is necessary to use such models, as the
lens-shaped self-assembled QDs with a quantum well confined to a
narrow wetting layer~\cite{Hawrylak96}, or, if the minor semiaxis
becomes comparable with the lattice constant, to proceed to models beyond the effective mass
approximation (see,e.g.\cite{Harper}).

\begin{figure}[t]
\includegraphics[width=0.45\textwidth]{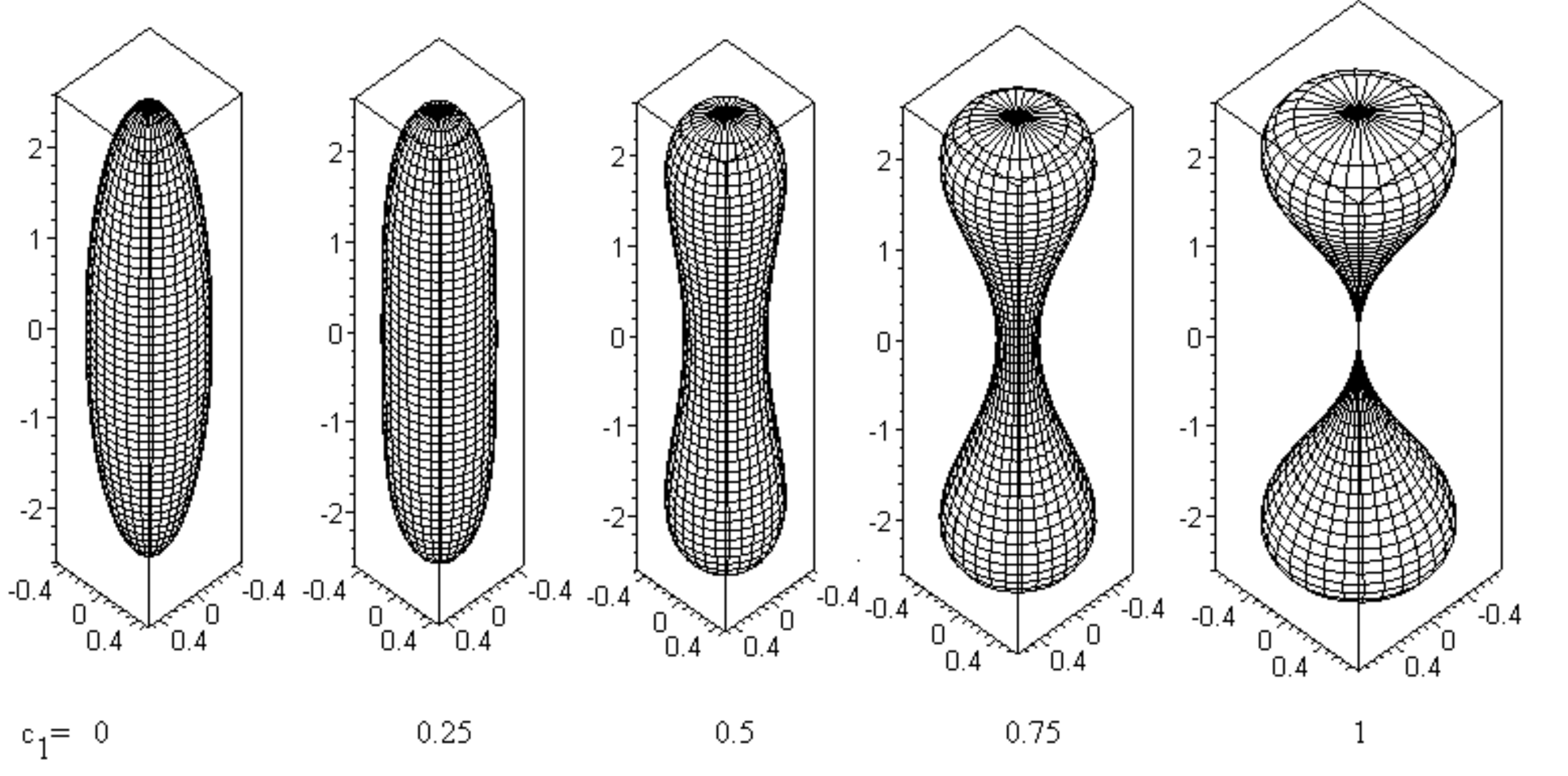}a
\includegraphics[width=0.45\textwidth]
{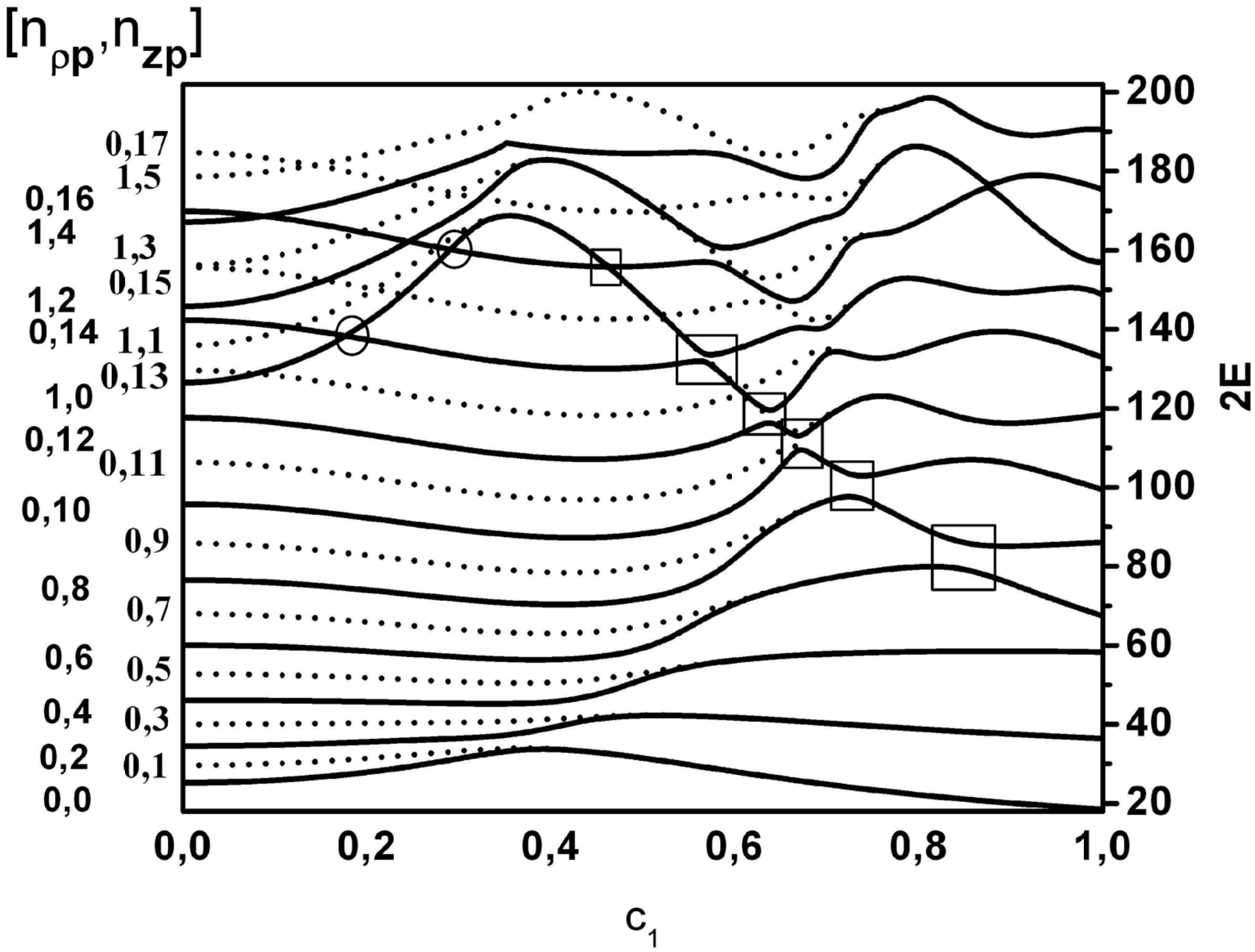}b

\caption
{a. The profile in plane $z,\rho$ of closed surface
generated by rotating of continuous curve $ \rho_{0} ( z;a,c,c_1)$
from (\ref{eq99a}) about z-axis  for $c=2.5$, $a=0.5$ vs
$c_1=0,0.25,0.5,0.75,1$. b. The energy levels of the
even and odd states of DSQD for model B  generated by (\ref{eq99a})
for $c=2.5$, $a=0.5$  vs $c_1$
 classified at $c_1=0$ by
adiabatic quantum numbers $n_{\rho p}$,  $n_{zp}$ and {\bf  $m=0$}
of PSQD. }
\label{gant051}
\end{figure}

\subsection{Models B for dumbbell QD}
For DQD at the fixed coordinate  $x_s$ of the slow subsystem the
motion of the particle
in the fast degree of freedom $x_{f}$
 is confined to a 2D potential
double well at $0\leq c_1\leq 1$ with the effective variable radius
\begin{eqnarray}   \label{eq99ab}
 \rho_{0} \left( x_{s}\right)\equiv \rho_{0} \left( x_{s};a,c,c_1 \right)
 =\frac{a}{c}\sqrt{c^2-x_s^2}\frac{x_s^2c_1^2+1-c_1^2}{c_1^2c^2/4+1-c_1^2}.
\end{eqnarray}
Fig. \ref{gant051} illustrates the transformation of the prolate
spheroidal shape of QD with $c=2.5$ and $a=0.5$, considered in the
previous Section, into a ``dumbbell''-type shape
and the corresponding evolution
of the lower part of the countable spectrum
$\tilde E(\zeta_{ac}=1/5, c_{1})/\tilde E_{R}$ = $2E_t$  of Model B
versus the deformation parameter $c_{1}$ at a few fixed values
$c_1=0,0.25,...,1$ from the interval $0\leq c_1\leq 1$. At
$c_1=0$ the discrete spectrum states are characterized by a set of
exact spheroidal or adiabatic cylindrical quantum numbers,
$\{n_{\xi},n_{\eta},m,\sigma\}$ or [$n_{\rho p},n_{zp},m,\sigma$].
Typically, one can see  exact crossing of energy levels having
different parity ($\sigma =\pm1$) with the growth of the deformation
parameter $c_{1}$, which  leads, first, to the quasidegeneracy of
these energy levels and then to their exact degeneracy at the
critical value $c_{1}=1$. On the other hand, for small values of the
deformation parameter $c_{1}$ one observes, first, exact crossings
(labelled with circles like in Fig.\ref{enektv}a above) of
similar-parity energy levels, replaced with the avoided crossings
(labelled with squares) for greater values of the deformation
parameter approaching the critical value $c_{1}=1$. A similar
picture was observed in the example of a 2D-Sinai billiard~\cite{Akishin97},
a 2D-quantum billiard with
the shape $x^{2}+y^{2}+\epsilon x^{3}=1$ and the deformation
parameter $\epsilon>0$, provided the so-called whispering gallery modes
and considered in~ \cite{Bayfield,BrunoCrespi1993},
as well as in the unidirectional far-field emission of coupled nonidentical microdisks ~\cite{avoid}.
\begin{figure}[t]
\includegraphics[width=0.8\textwidth,height=0.8\textwidth]{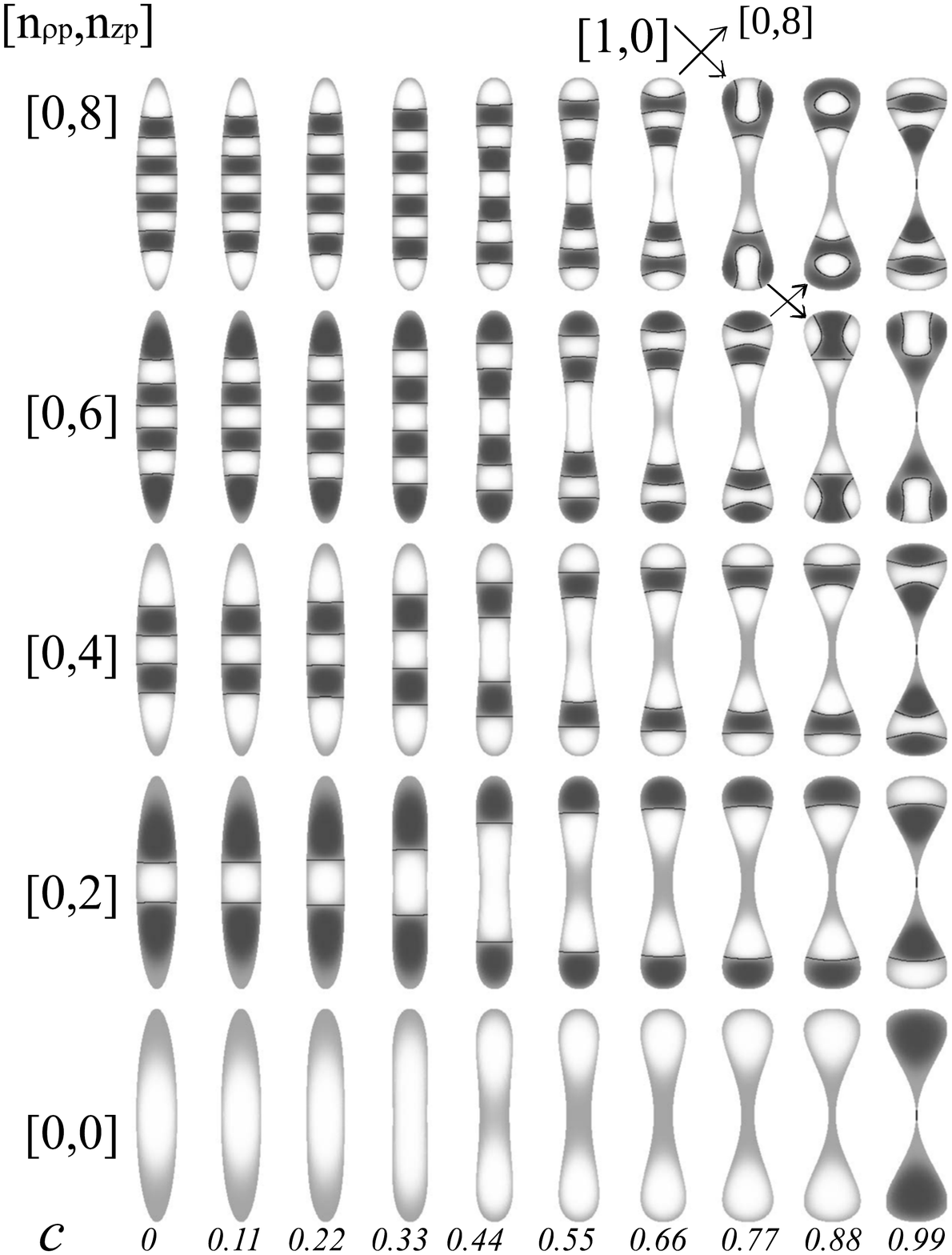}
\caption {
Contour lines of the first five eigenfunctions of model B of DSQD at  $a=2.5$, $c=0.5$ and several values of  $c_1$. Light and dark  inflections are positive and  negative values of eigenfunctions and lines are  eigenfunction  nodes. The adiabatic cylindrical quantum numbers  [$n_{\rho p}$, $n_{zp}$] are given at $m=0$. Crossing arrows mean the transformation of nodes of pair of eigenfunctions after passing value of parameter in which avoided crossing of corresponding pair of eigenvalues was taking place in Fig. \ref{gant051}.b.
}  \label{gant052}
\end{figure}

In Fig. \ref{gant052} we show the evolution of the first
five eigenfunctions with the increasing deformation parameter
values $c_1=c_1=0,0.11,...,0.99$.

The transformation of eigenfunctions when passing the avoided
crossing points (labelled with squares) in Fig. \ref{gant051} b,
is shown in Fig. \ref{gant052} for model B of DQD (labelled with
arrows). Comparing these Figures, one can see that if the
eigenfunctions are ordered in accordance with the increasing
eigenvalues of the BVPs, then the number of nodes is not invariant
under the variation of the parameter $c_{1}$ from $c_{1}=0$ to
$c_{1}=1$ in the potentials (\ref{eq99ab}).
In particular, in Fig. \ref{gant052} one can see that the
eigenfunction of the state $[n_{\rho p}=0,n_{zp}=6,m=0,\sigma=+1]$
at $c_1=0.99$ has the same number of nodes as the eigenfunction of
the state $[n_{\rho p}=1,n_{zp}=0,m=0,\sigma=+1]$  at $c_1=0$.
Above we could already observe this in Fig.\ref{awe} at $a=1$
(up-going arrow) after several exact and avoided crossings of the
corresponding energy levels in Fig. \ref{gant051}b). At the same
time, the eigenfunction of the state $[n_{\rho
p}=0,n_{zp}=8,m=0,\sigma=+1]$ at $c_1=0.99$ after avoided crossing
of the corresponding energy levels in Fig. \ref{gant051}b) has the
same number of nodes as the eigenfunction of the state $[n_{\rho
p}=6,n_{zp}=0,m=0,\sigma=+1]$ at $c_1=0$.

\section{
{Absorption Coefficient} for an Ensemble of QDs}\label{5} One can
use the mentioned differences in the energy spectra to verify the
considered models of QDs by calculating the absorption coefficient
$K(\omega^{ph},\tilde a,\tilde c,)$ of an ensemble of identical
semiconductor QDs \cite{Efros1982}:
\begin{eqnarray}&&\label{coa1}
\tilde K(\tilde\omega^{ph},\tilde a,\tilde c)= \sum_{\nu,\nu'}\tilde
K_{\nu,\nu'}(\tilde\omega^{ph},\tilde a,\tilde c)=\tilde
A\sum_{\nu,\nu'} \tilde I_{\nu,\nu'} \delta(\hbar
\tilde\omega^{ph}-\tilde W_{\nu\nu'}),
\\&&\nonumber
\tilde I_{\nu,\nu'}=|\int\tilde \Psi^{e}_{\nu}(\tilde {\vec r};\tilde
a,\tilde c,)\tilde \Psi^{h}_{\nu'}((\tilde {\vec r};\tilde a,\tilde c,))d
\tilde {\vec r}|^{2},  \,\,\,\,
\tilde W_{\nu\nu'}=\tilde E_{g}+\tilde E^{e}_{\nu}(\tilde a,\tilde c)
+\tilde E^{h}_{\nu'}(\tilde a,\tilde c),
\end{eqnarray}
where $\tilde A$ is proportional to the square of the matrix
element in the Bloch decomposition, $\tilde \Psi^{e}_{\nu}(u)$ and
$\tilde \Psi^{h}_{\nu'}$ are the eigenfunctions of an electron
($e$) and a heavy hole ($h$), $\tilde E^{e}_{\nu}$ and $\tilde
E^{h}_{\nu'}$ are the energy eigenvalues for an electron ($e$) and
a heavy hole ($h$), depending on the semiaxis size $\tilde c,
\tilde a$ for OSQD (or $\tilde a,\tilde c$ for PSQD) and the
adiabatic set of quantum numbers $\nu=[n_{zo},n_{\rho o},m]$ and
$\nu'=[n_{zo}',n_{\rho o'},m']$ ($\nu=[n_{\rho p},n_{zp},m]$ and
$\nu'=[n_{\rho p}',n_{z p}',m']$), where $m'=-m$, $\tilde E_{g}$
is the band gap width in the bulk semiconductor, $\tilde
\omega^{ph}$ is the incident light frequency, $\tilde W_{\nu\nu'}$
is the inter-band transition energy for which $\tilde
K(\tilde\omega^{ph})$ has the maximal value. We rewrite the
expression (\ref{coa1}) using  dimensionless quantities in reduced
atomic units
$$
\tilde K(\omega^{ph},\tilde a,\tilde c)= \tilde A \tilde
E_{g}^{-1}\sum_{\nu,\nu'} \tilde I_{\nu,\nu'}
\delta[f_{\nu,\nu'}(u)], $$ $$ f_{\nu,\nu'}(u)=
\lambda_{1}-(2E_{g})^{-1}(2 E^{e}_{\nu}(a, c)+2
E^{h}_{\nu'}(a,c)(\mu_{h}/\mu_{e})),
$$
where the parameter will $u$ be defined below, $\lambda_1=(\hbar
\tilde \omega^{ph} - \tilde E_{g})/\tilde E_{g}$ is the energy of
the optical interband transitions scaled to $\tilde E_{g}$,
$2E_{g}= \tilde E_{g}/\tilde E^{e}_{R}=1.43/(5.27\cdot10^{-3})$ is
the dimensionless band gap width. {For both electron and hole
carriers the dimensionless energies $2E_{\nu}^{e}= \tilde
E_{\nu}^{e}/\tilde E^{e}_{R}$ and
$2E_{\nu}^{h}(\mu_{h}/\mu_{e})=\tilde E_{\nu}^{h}/\tilde
E^{e}_{R}$ are expressed in the same reduced atomic units $\tilde
E^{e}_{R}$.}

Now consider an ensemble of OSQDs (or PSQDs) with different values
of the minor semiaxis $c=u_{o}\bar c$ (or $a=u_{p}\bar a$) determined by
the random parameter $u=u_{o}$ (or $ u=u_{p}$). The corresponding
minor semiaxis mean value is  $\bar c$ at fixed major semiaxis $a$ (or $\bar
a$ at fixed major semiaxis $c$) and the appropriate distribution
function is  $P(u_{o})$ (or $P(u_{p})$). Conventionally, they use
the normalized Lifshits-Slezov
 $P(u)\equiv P^{LS}(u)$ \cite{LS1958} or Gaussian
$P(u)\equiv P^{G}(u)$ distribution functions ($\int P(u)du=\int u
P(u)du=1$):
$$
P^{LS}(u):=\left\{\begin{array}{ll}\frac{3^4eu^2\exp(-1/(1-2u/3))}{2^{5/3}(u+3)^{7/3}(3/2-u)^{11/3}}, &u\in (0,3/2);\\
 0,& \mbox{otherwise}\end{array}\right.
$$
$$
P^{G}(u):=1/\sqrt{2\pi}/\sigma\exp(-(u-1)^2/(2\sigma^2)),
$$
where  $\bar u= \int u P^{G}(u)du=1$ is the mean value of $u$ and
$\sigma^{2}=(\int (u-\bar u)^{2} P^{G}(u)du)$ is the variance. The
absorption coefficient of an ensemble of semiconductor QDs with
different dimensions of minor semiaxes is then expressed as
$$
\tilde K^{o}(\omega^{ph},\bar {\tilde a}, \tilde c)
=\int \tilde K(\omega^{ph}, \bar {\tilde a}, \tilde c, u_{o})P(u_{o})du_{o}, $$ $$
\tilde K^{p}(\omega^{ph},\tilde a,
\bar {\tilde c})=\int \tilde K(\omega^{ph},   \tilde a, \bar {\tilde c}, u_{p})P(u_{p})du_{p}.
$$
Taking the known properties of the $\delta$-function into account,
we arrive at the analytical expression for the the absorption
coefficient $\tilde K(\omega^{ph},\tilde a,\tilde c)$ of a system
of semiconductor QDs with a distribution of minor semiaxes:
\begin{equation}\label{kk}
\frac{\tilde K(\omega^{ph})}{\tilde K_{0}}=
\sum_{\nu,\nu',s} \frac{\tilde K_ {\nu,\nu'}(\omega^{ph})}{\tilde K_{0}},
\quad \frac{\tilde K_ {\nu,\nu'}(\omega^{ph})}{\tilde K_{0}}=
\tilde I_ {\nu,\nu'}
\left|\left.\frac {df_{\nu,\nu'}(u)}{du}\right|_{u=u_s}\right|^{-1}  P\left( u_s \right),
\end{equation}
where $\tilde K_{0}= \tilde A^{-1}\tilde E_{g}$ is the
normalization factor, $u_s$ are the roots of the equation
$f_{\nu,\nu'}(u_s)=0$.

In particular, for Model B of OSQD or PSQD we have the interband
overlap $\tilde I_{\nu,\nu'}=\delta_{n_{\rho o},n_{\rho
o}'}\delta_{n_{zo},n_{zo}'}\delta_{m,-m'}$ for OSQD, \\ $\tilde
I_{\nu,\nu'}=(J_{1+|m|}(\alpha_{n_{\rho
p}+1,|m|})/J_{1-|m|}(\alpha_{n_{\rho p}+1,|m|}))^2
\delta_{n_{zp},n_{zp}'}\delta_{n_{\rho p},n_{\rho p}'}
\delta_{m,-m'}$ for PSQD, and the selection rules
$n_{zo}=n_{zo}'$, $n_{\rho o}=n_{\rho o}'$, and $m=-m'$ or
$n_{\rho p}=n_{\rho p}'$, $n_{zp}=n_{zp}'$ and $m=-m'$,
respectively. Note that the contributions of non-diagonal matrix
elements to the energy values are about 1\% for OSQD and PSQD of
Model B; then in the Born-Oppenheimer approximation of the order
$b_{max}$  for the absorption coefficient we get
\begin{eqnarray}
\label{ca01} f_{\nu,\nu'}(u)=\lambda_{1}-\sum_{j=0}^{b_{max}} \check
E^{(j)}u^{j-2}. \end{eqnarray}
Here the coefficients $\check E^{(j)}$ are defined by
\begin{eqnarray} && \nonumber
\check E^{(j)}= (2E_{g})^{-1}E^{(j)}_{io}
\omega_{\rho;n_o}^{2-j}(\bar c)(1+\mu_{e}/\mu_{h})
\\&& \mbox{or} \quad \check E^{(j)}=(2E_{g})^{-1}E^{(j)}_{ip}
\omega_{z;n_{\rho p}}^{2-j}(\bar a)(1+\mu_{e}/\mu_{h}).
\\ &&
\label{ca01}
\omega_{\rho;n_{o}}(\bar c)=\pi n_{o}/(a\bar c) ,
\quad \omega_{z;n_{\rho p}}(\bar a)=\alpha_{n_{\rho p}+1,|m|}/(\bar
ac).
\\ &&\nonumber
E^{(0)}_{io}= a^2/4,\quad E^{(1)}_{io}=(2n_{\rho o}\!+\!{|m|}\!+\!1)
,\\&& E^{(2)}_{io}=(6n_{\rho o}{|m|}\!+\!2\!+\!6n_{\rho
o}\!+\!6n_{\rho o}^2\!+\!{|m|}^2\! +\!3{|m|})a^{\!-\!2},
\\&& \nonumber E^{(3)}_{io}=\!3(6n_{\rho o}\!+\!3{|m|}\!+\!2\!+\!{|m|}^2\!+\!6n_{\rho o}^2\!
+\!6n_{\rho o}{|m|}\!+\!4n_{\rho o}^3\\&& ~~~~+\!6{|m|}n_{\rho o}^2\!
+\!2{|m|}^2n_{\rho o})a^{\!-\!4}/2,
\\&& \nonumber E^{(0)}_{ip}=
c^2,\quad E^{(1)}_{ip}=(2n_{zp}\!+\!1),
\quad E^{(2)}_{ip}=\!+\!3(2n_{zp}\!+\!2n_{zp}^2\!+\!1)c^{-2}/4,
\\  &&\nonumber E^{(3)}_{ip}=\!3(3n_{zp}^2\!+\!7n_{zp}\!+\!2n_{zp}^3\!
+\!3)c^{-4}/16.
\end{eqnarray}
The coefficients of the order $b_{\max}\ge4$ are calculated by the
perturbation theory algorithms \cite{CASC07,CASC09} using exact solutions of 2D and 1D oscillators with \textit{adiabatic} frequencies $\omega_{\rho;n_{o}}(\bar c)$ and $\omega_{z;n_{\rho p}}(\bar a)$ from (\ref{ca01}) that distinguish from conventional ones, for example, $\omega_\rho$ and $\omega_z$ using in section \ref{3}.1 or in \cite{Rassey58,Poenaru08}.
{The accuracy of such approximations up to $b_{\max}=5$ is about 4
-- 6 decimal digits in comparison with the numerical results of
the crude diagonal adiabatic approximation (CDAA) of
Eqs.(\ref{sp23}) without $H_{ii}(x_{s})$  for the states from Fig.
\ref{enekts}a at $c=0.5$ and Fig. \ref{enektv}a at $a=0.5$.  In
the case  $a=c=1$ the accuracy is  only about two decimal digits
in comparison with the CDAA of the exact spectrum Eq.
(\ref{exact}) of model B of SQDs~\cite{Efros1982}.}

Note that in model B  $2E_{io}$ and $2E_{ip}$  monotonically depend
upon the parameter $u$ and, therefore, the algebraic equation
$f_{\nu,\nu'}(u)=0$ has the only solution in the considered domain
of definition. Using the notations $\lambda_1'=\lambda_1$ for
$b_{max}=1$ and $\lambda_1'=\lambda_1-E^{(2)}_{io}$, or
$\lambda_1'=\lambda_1-E^{(2)}_{ip}$ for $b_{max}\ge 2$, we rewrite
this equation in the Born-Oppenheimer approximations up to the third
order $b_{max}\leq 3$
$$f_{\nu,\nu}(u)=\lambda'_{1}- \check E^{(0)}u^{-2}- \check E^{(1)}u^{-1}- \check E^{(3)}u=0,$$
which has the required roots $u_{1}=u_{1}^{(b_{max})}$:
$$
u_{1}^{(1,2)} = (2\lambda_1')^{-1}(\check E^{(1)}+((\check E^{(1)})^2
+4\lambda_1'\check E^{(0)})^{1/2})),
$$ $$
u_{1}^{(3)} =u^{(2)}+\check E^{(3)}(u_{1}^{(2)})^4/(2\check E^{(0)}+\check E^{(1)}u_{1}^{(2)}).
$$
\begin{figure}[t]
\includegraphics[width=0.3\textwidth,height=0.3\textwidth]{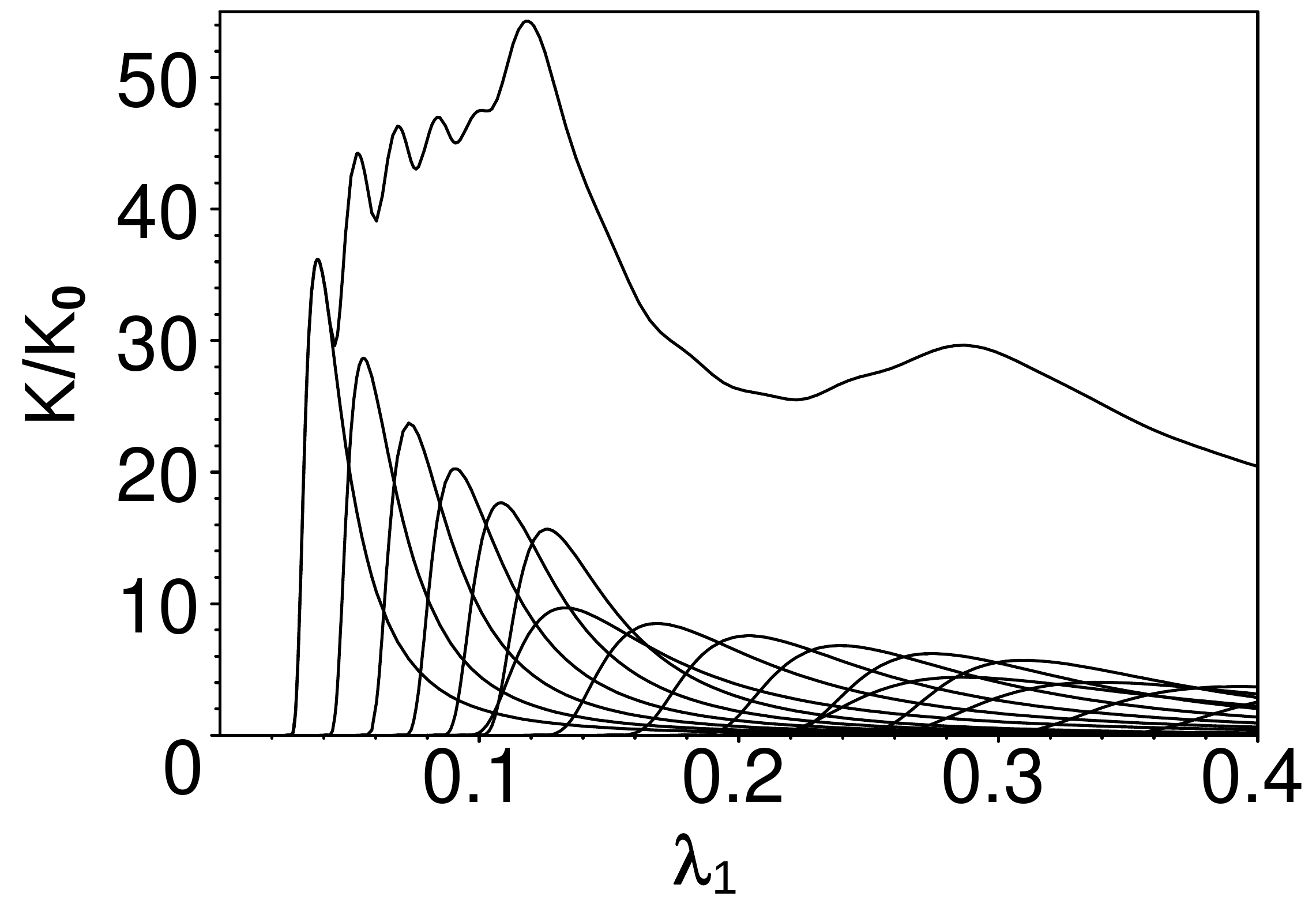}   
\includegraphics[width=0.3\textwidth,height=0.3\textwidth]{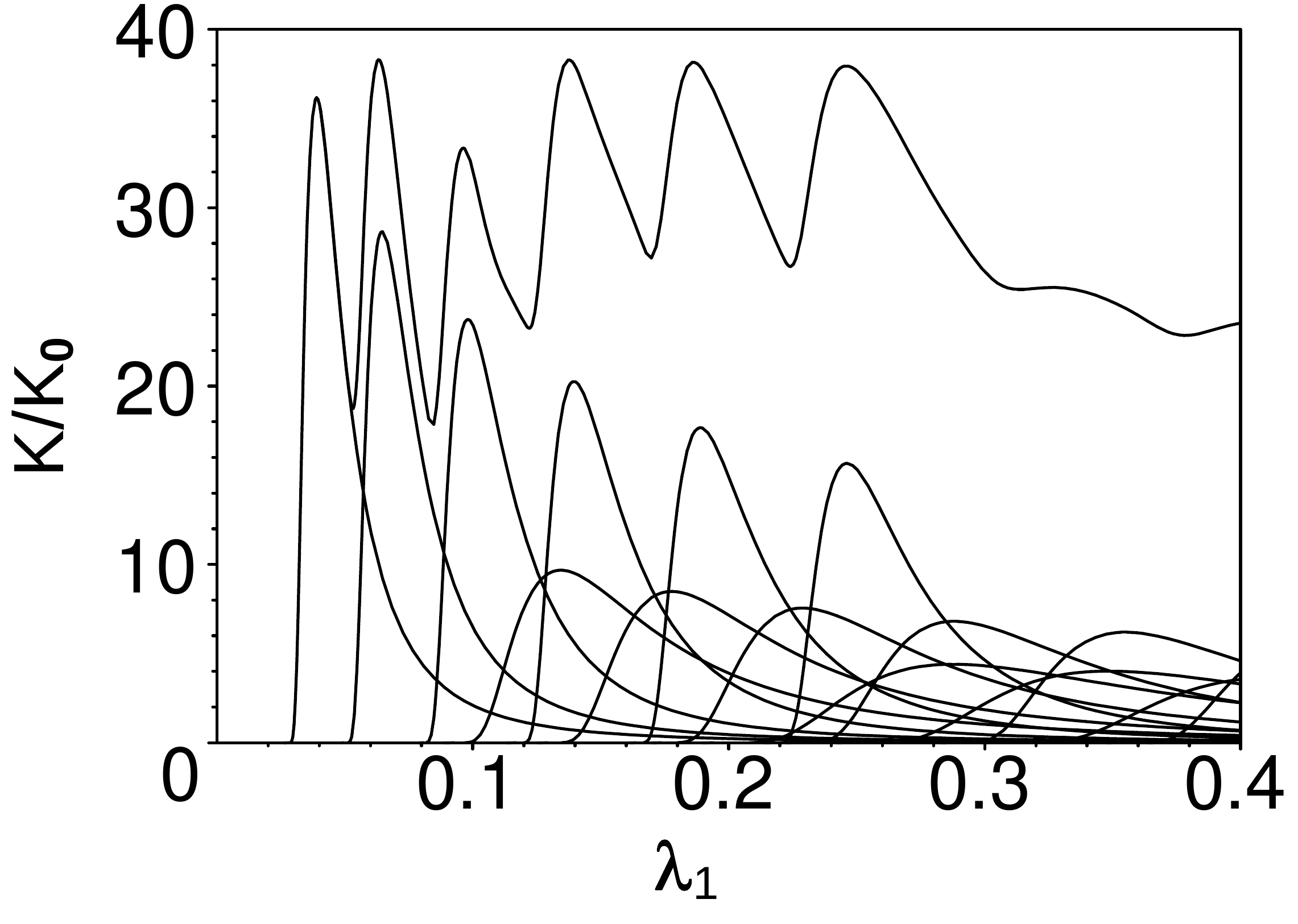}
\includegraphics[width=0.3\textwidth,height=0.3\textwidth]{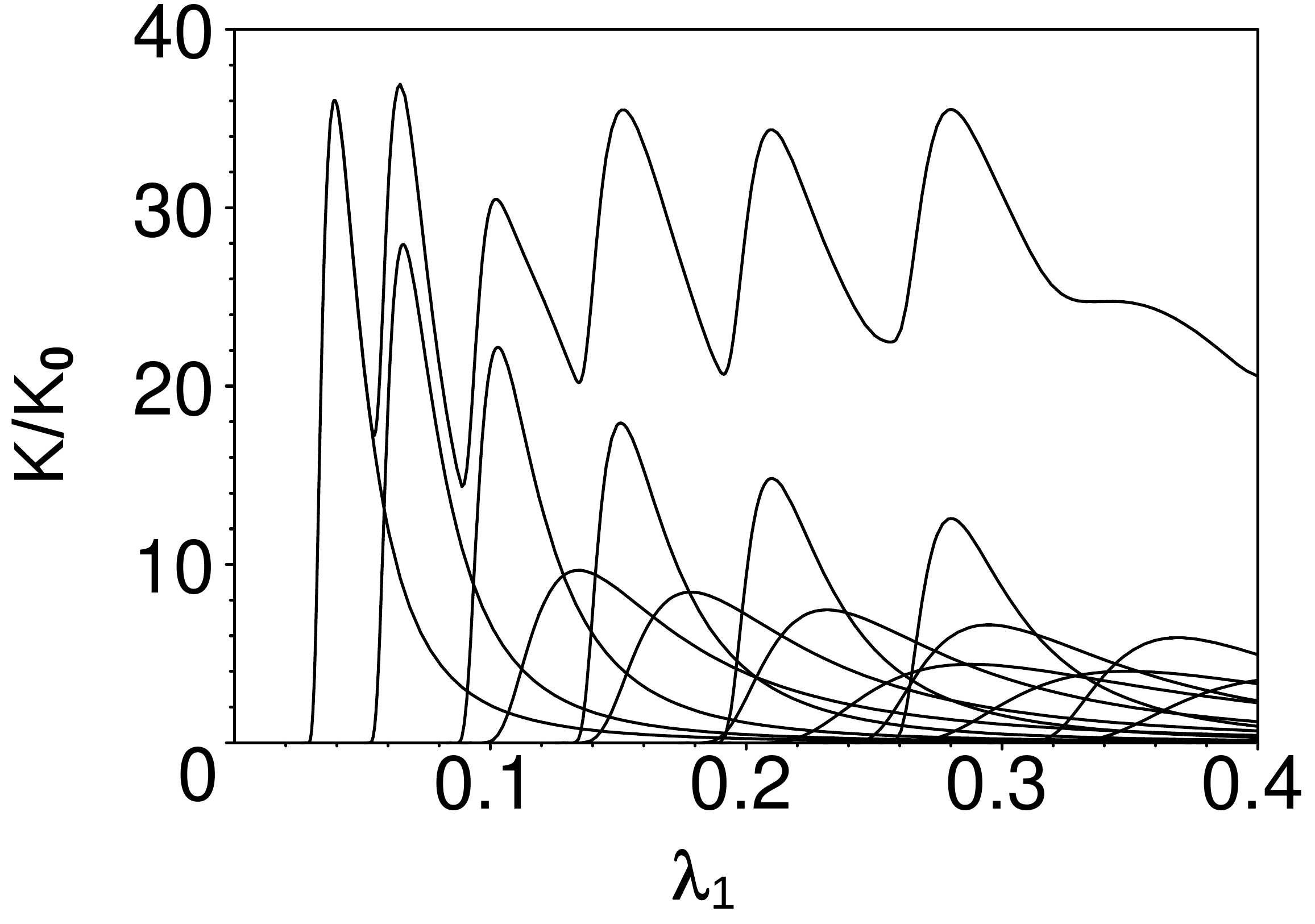}\\
\includegraphics[width=0.3\textwidth,height=0.3\textwidth]{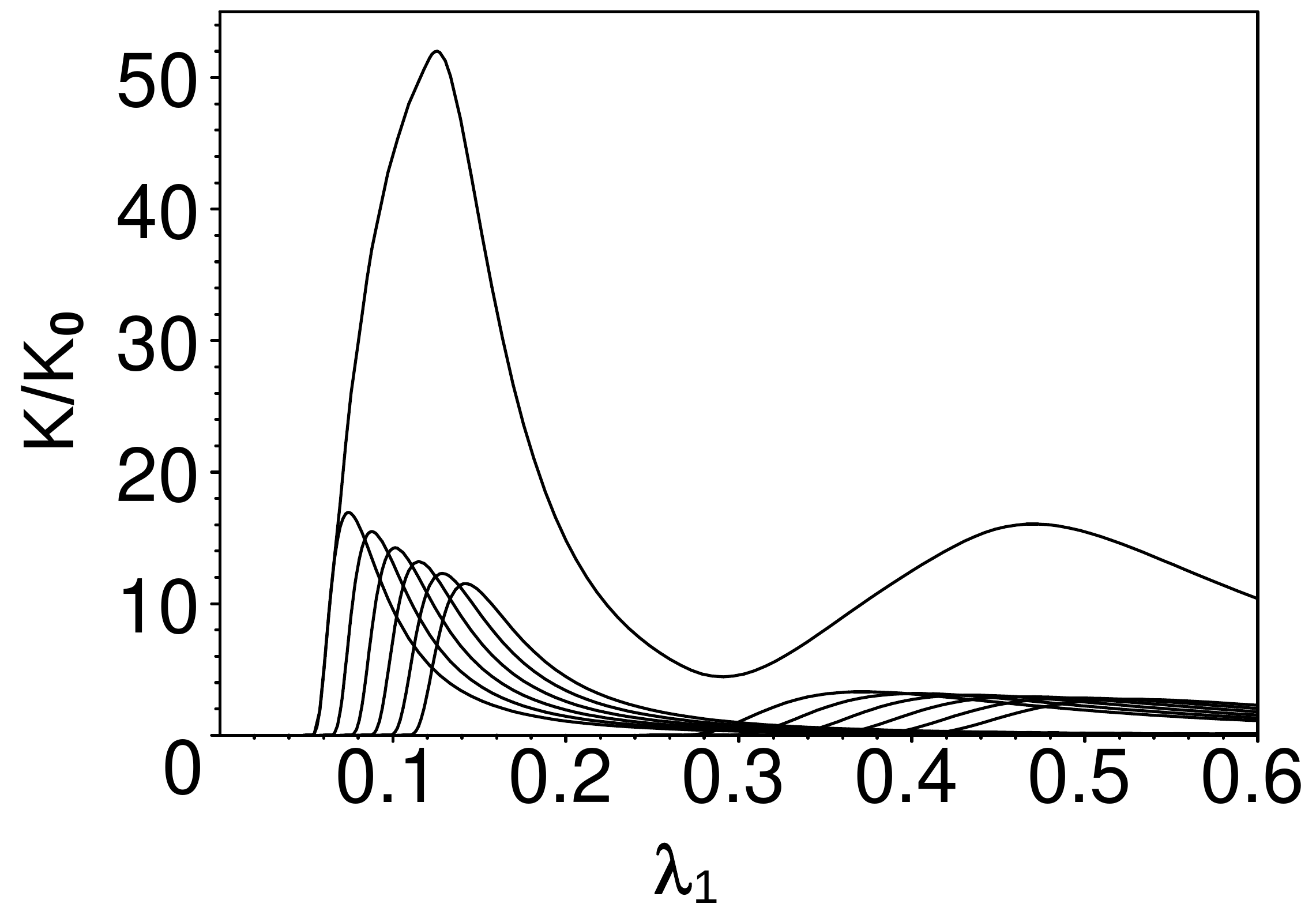} 
\includegraphics[width=0.3\textwidth,height=0.3\textwidth]{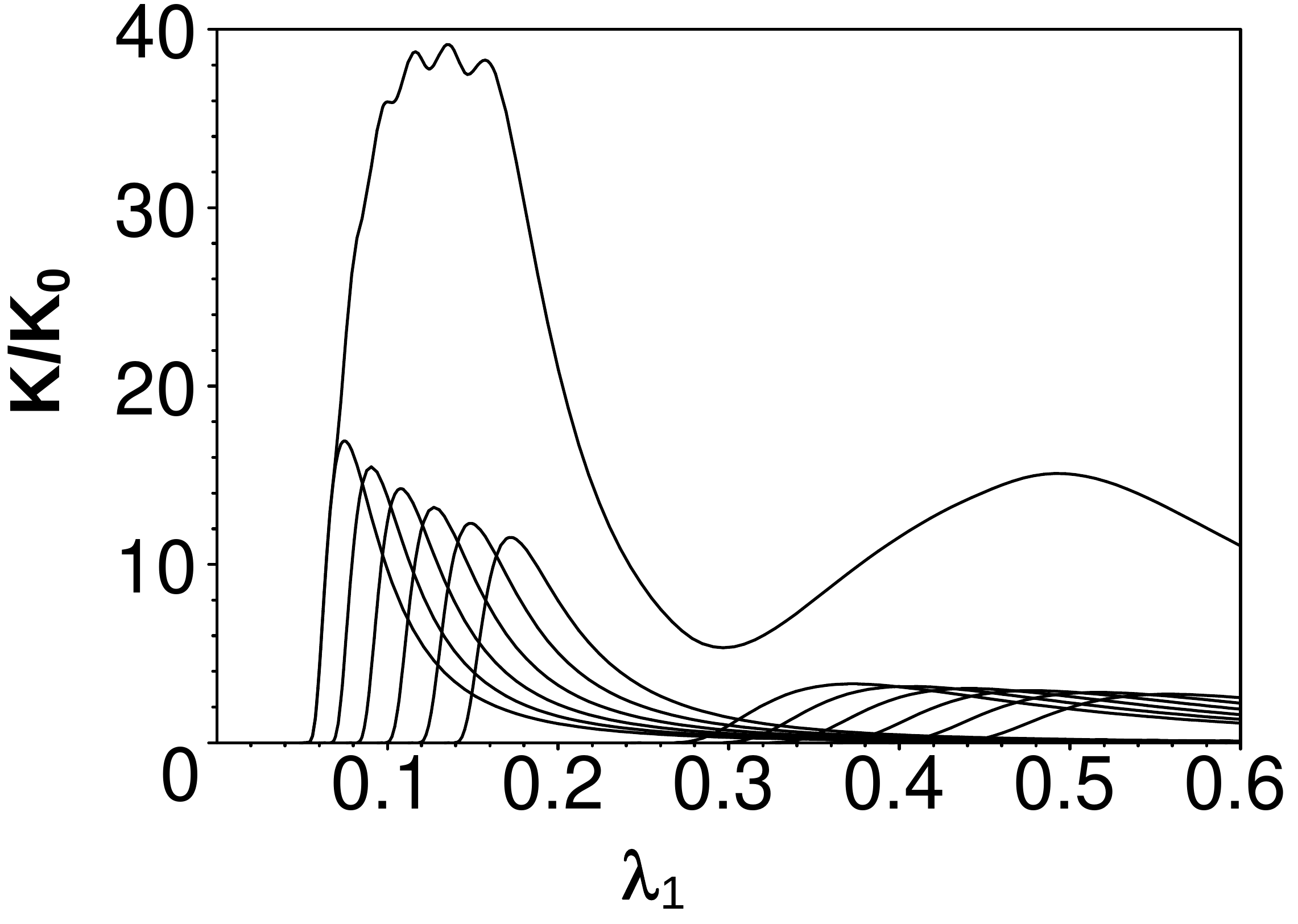}
\includegraphics[width=0.3\textwidth,height=0.3\textwidth]{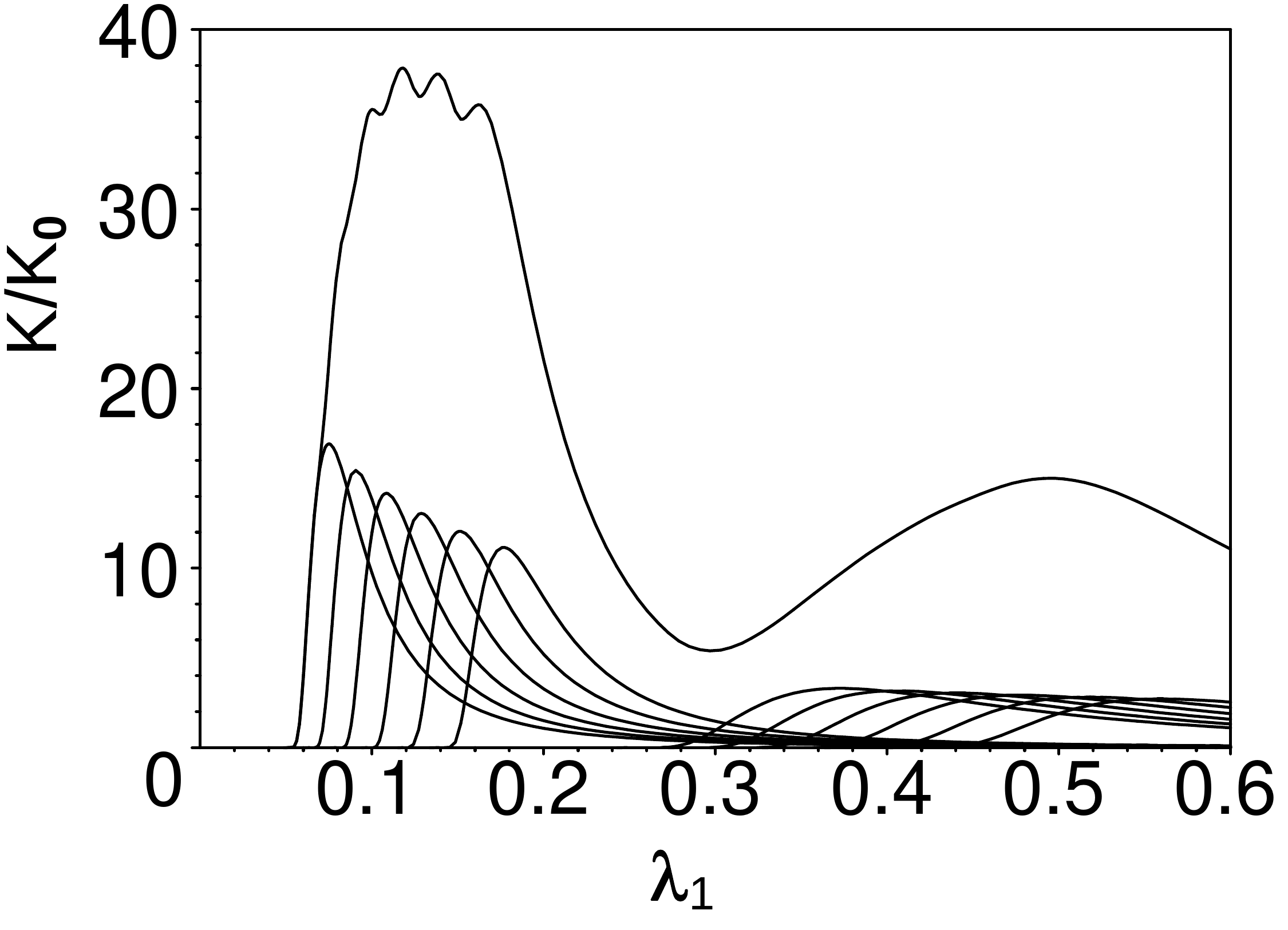}\\
\caption{Absorption coefficient $K/K_{0}$ from (\ref{kk})
consists of sum of the first
partial contributions vs the energy $\lambda=\lambda_1$
of the optic interband transitions for the Lifshits-Slezov distribution
in first, second and third (from left to right) Born Oppenheimer approximations:
(top panels) for assemble of OSQDs  $\bar c=0.5$, $a=2.5$
(summation by $n_o=1,2,3$, $n_{\rho o}=0,1,2,3,4,5$, $m=0$),
(bottom panels)
for assemble of PSQDs  $\bar a=0.5$, $c=2.5$
(summation by $n_{p}=1,2,3$, $n_{zp}=0,1,2,3,4,5$, $m=0$).
} \label{k8}
\end{figure}

For the Lifshits-Slezov distribution Fig. \ref{k8} displays the
total absorption coefficients $ {\tilde K(\omega^{ph})}/{\tilde
K_{0}}$ and the partial absorption coefficients  ${\tilde K_
{\nu,\nu}(\omega^{ph})}/{\tilde K_{0}}$, that form the corresponding
partial sum (\ref{kk}) over a fixed set of quantum numbers $\nu$ at
$m=-m'=0$. One can see that the summation over the quantum numbers
$n_{zo}$ (or $n_{\rho p}$) numerating the nodes of the wave function
with respect to the fast variable gives the corresponding main
maxima of the total absorption coefficients for the ensemble of QDs
with distributed dimensions of minor semiaxis, while the summation
over the quantum number $n_{\rho o}$ (or $n_{zp}$) that label the
nodes of the wave function with respect to the slow variable leads
to the increase of amplitudes of these maxima and to appearing
secondary maxima in the case of sparer energy levels of Model B
OSQDs (or PSQDs)

\begin{figure}[t]
\includegraphics[width=0.44\textwidth,height=0.44\textwidth]{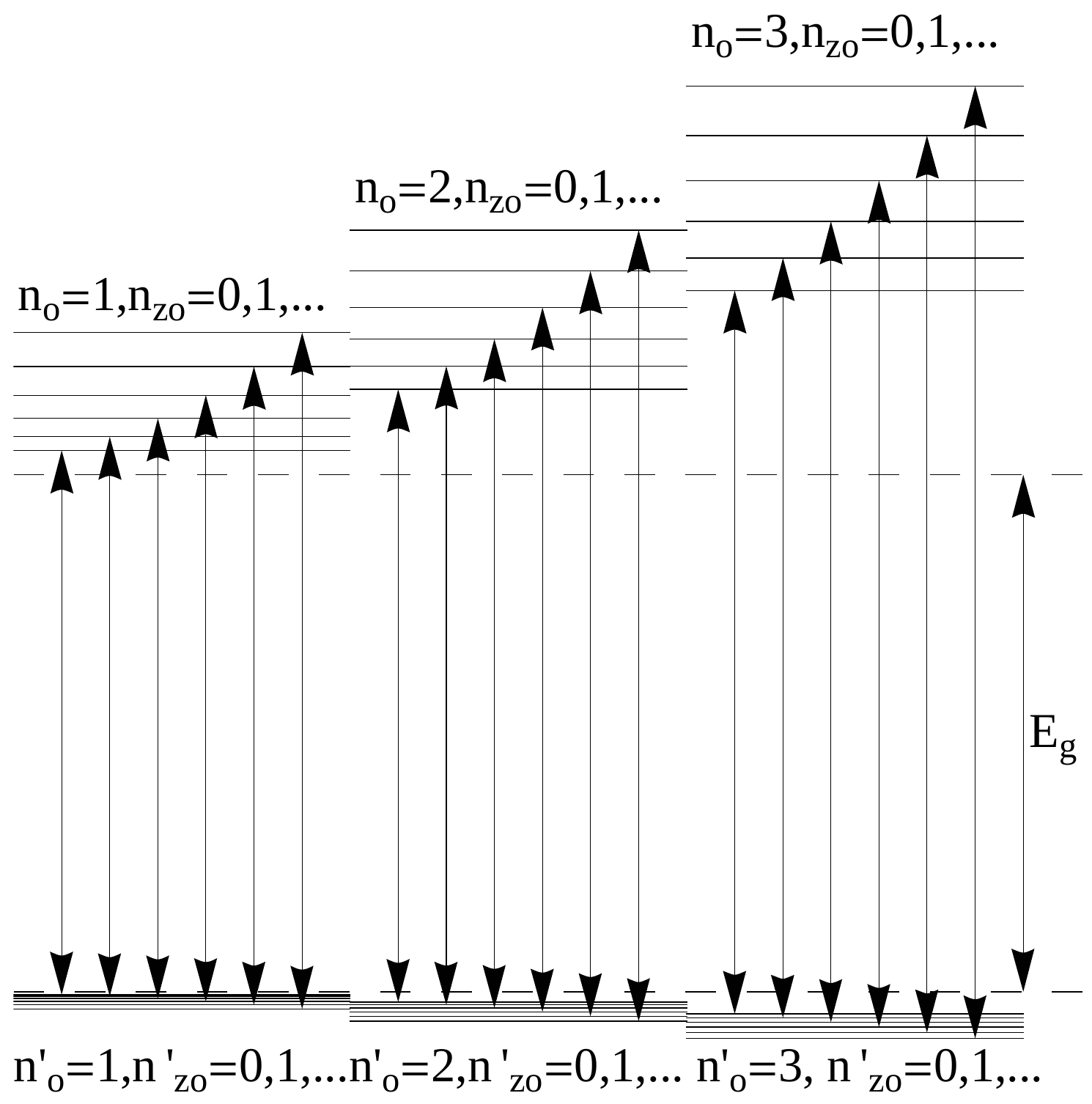}
\includegraphics[width=0.44\textwidth,height=0.44\textwidth]{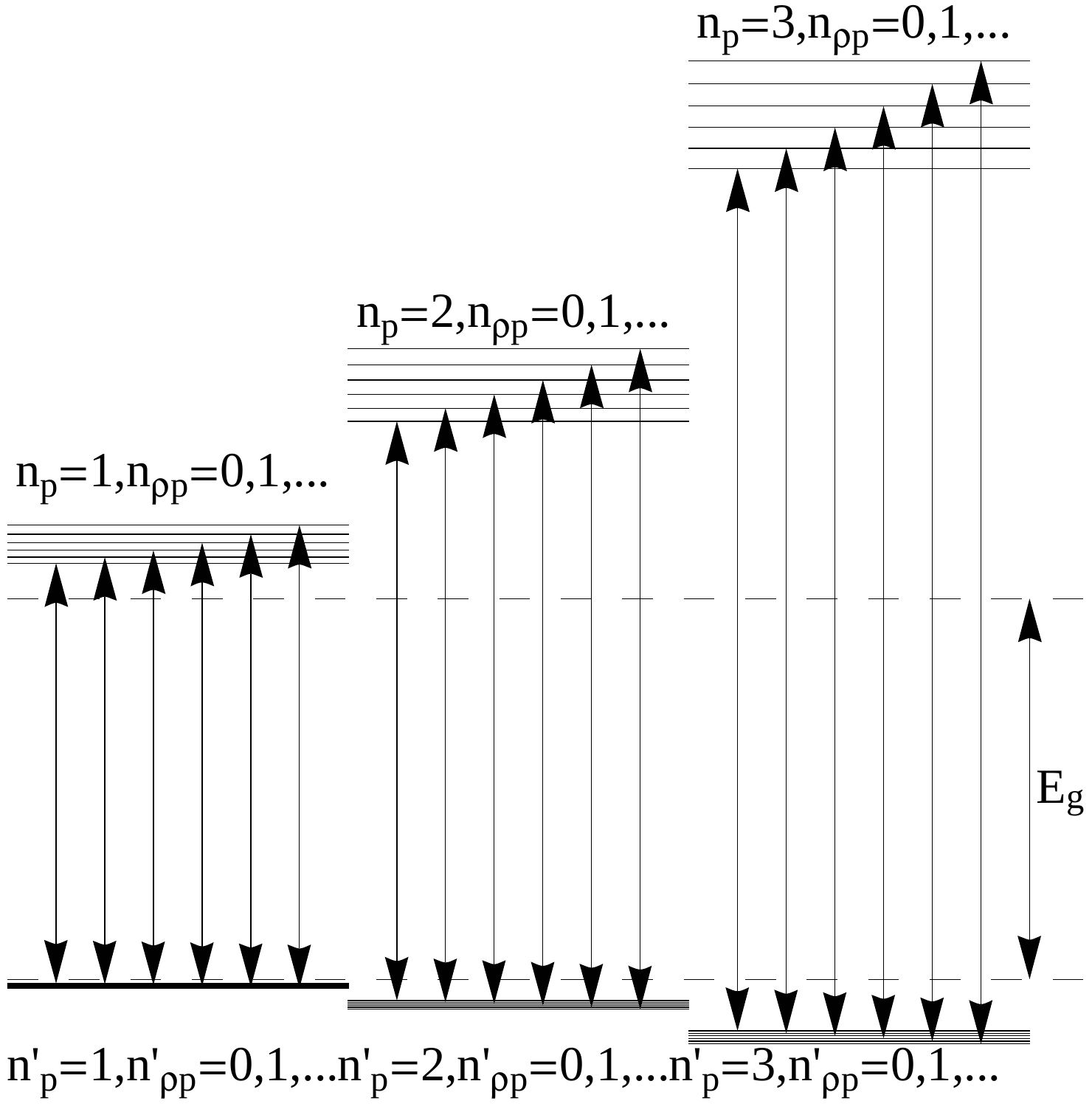}
\caption{Schematic plots of interband transition corresponded to  Fig. \ref{k8}.
} \label{efros}
\end{figure}

In the regime of strong dimensional quantization the frequencies of
the interband transitions between the levels  $n_{o}=1, n_{\rho
o}=0, m=0$ for OSQD or $n_{p}=1,n_{z p}=0, m=0$ for PSQD
{in the BO1, at the fixed values $\tilde a = 2.5 a_{e}$ and
$\tilde c = 0.5 a_{e}$ for OSQD or $\tilde a = 0.5 a_{e}$ and
$\tilde c = 2.5 a_{e}$ for PSQD, are equal to
$\tilde \omega^{ph}_{100}=2.17 \times 10^{13} $ s$^{-1}$ or
$\tilde \omega^{ph}_{100}=3.32 \times 10^{13} $ s$^{-1}$
($\tilde \omega^{ph}_{100}= \hbar^{-1}\tilde W_{100,100}$ with the
accuracy to $3\%$ and $0.5\%$, respectively),  corresponding to
the
infrared spectral region~\cite{79,79a}.} With decreasing semiaxis
the threshold energy increases, because the ``effective'' band gap
width increases, which is a consequence of the enhancement of
dimensional quantization. Therefore, the above frequency is
greater for PSQD than  for OSQD, because the SQ implemented in two
direction of the plane (x,y) is effectively greater than that in
the direction of the $z$ axis solely at similar values of
semiaxes. Higher-accuracy calculations reveal an essential
difference in the frequency behavior of the absorption coefficient
for interband transitions (see Fig. \ref{efros}) in systems of semiconductor OSQDs or
PSQDs having a distribution of minor semiaxes, which can be used
to verify the above models.

\section{Conclusions}
The presented examples of the analysis of energy spectra of SQD,
OSQD, PSQD, and DQD models with three types of axially symmetric
potentials demonstrate the efficiency of the developed
computational scheme and SNA. Only Model A (anisotropic harmonic
oscillator potential) is shown to have an equidistant spectrum,
while Models B and C (wells with infinite and finite walls height)
possess non-equidistant spectra. In Model C, there is a finite
number of energy levels. This number becomes smaller as the
parameter $a$ or $c$ ($\zeta_{ac}$ or $\zeta_{ca}$) is reduced
because the potential curve (lower bound) moves into the
continuum. Models A and B have countable discrete spectra. This
difference in spectra allows verification of SQD, OSQD, and PSQD
models using the experimental data \cite{Gambaryan}, e.g.,
photoabsorption, from which not only the energy level spacing, but
also the mean geometric dimensions of QD may be derived
\cite{79,Trani,Lepadatu}. { The considered examples of calculating
the absorption coefficient for ensembles of OSQDs or PSQD's with
random minor semi-axes in model B have proved the possibility of a
similar verification.} It is shown that there are critical values
of the ellipsoid aspect ratio, at which in the approximation of
effective mass the discrete spectrum of the models with
finite-wall potentials turns into a continuous one. Hence, using
the experimental data, it is possible to verify different QD
models like the lens-shaped self-assembled QDs with a quantum well
confined to a narrow wetting layer~\cite{Hawrylak96}, or to
determine the validity domain of the effective mass approximation,
if a minor semiaxis becomes comparable with the lattice constant
and to proceed opportunely to more adequate models such as
\cite{Harper}.

Further development of the method, {symbolic-numerical
algorithms}, and the software package is planned for solving the
quasi-2D and quasi-1D BVPs with both discrete and continuous
spectrum, which are necessary for calculating the optical
transition rates, channeling and transport characteristics in the
models like quantum wells or quantum wires and low-energy barrier
nuclear reactions.

The authors thank Profs. V.P. Gerdt and   V.A. Rostovtsev for collaboration and Profs. V. I. Furman, L.G. Mardoyan,
G.S. Pogosyan  for useful discussions. This work
was done within the framework of the Protocols No. 3967-3-6-09/11
and 4038-3-6-10/13 of collaboration between JINR (Dubna), RAU
(Erevan) and SSU (Saratov) in dynamics of low dimensional quantum
models and nanostructures in external fields. The work was supported
partially by RFBR (grants 10-01-00200 and 11-01-00523), and by the
grant No. MK-2344.2010.2 of the President of Russian Federation.


\begin{thebibliography}{99}
\bibitem{Harrison}  P. Harrison, \emph{Quantum Well, Wires and Dots}
(Wiley, New York, 2005).
\bibitem{Gambaryan}  K.M. Gambaryan,
Nanoscale Res Lett., DOI 10.1007/s11671-009-9510-8 (2009)
\bibitem{Hayrutyunyam} V. A. Harutyunyan et al, Phys. E \textbf{36}, 114 (2007).
\bibitem{Hawrylak96}  A. Wojs et al, 
Phys. Rev. B  \textbf{54}, 5604  (1996).
\bibitem{Hayk} L.A. Juharyan et al, 
Solid State Comm. \textbf{139}, 537 (2006).
\bibitem{79} K.G. Dvoyan et  al,
Nanoscale Res. Lett. \textbf{2}, 601 (2007).
\bibitem{79a} K.G. Dvoyan et  al,
Nanoscale Res. Lett. \textbf{4}, 106 (2009);
Proc. SPIE \textbf{7998}, 79981F (2010).
\bibitem{KimZubarev}  Y. E. Kim and A. L. Zubarev,
 Phys. Lett. A \textbf{289}, 155 (2001).
\bibitem{CNI2000} G.  Cantele et al, 
J. Phys. Condens. Matt.  \textbf{12}, 9019 (2000).
\bibitem{Trani} F. Trani et al, 
 Phys. Rev. B \textbf{72}, 075423 (2005).
\bibitem{Lepadatu} A.-M. Lepadatu et al,
J. Appl. Phys. \textbf{107}, 033721 (2010).
\bibitem{spherqd} A. Bagga, P. K. Chattopadhyay, and S. Ghosh,
arXiv: cond-mat/0406517v1 (2004).
\bibitem{Filikhin} I. Filikhin et al. Physica E \textbf{41} (2009) 1358--1363
\bibitem{mobius} J. Gravesen and M. Willatzen
Phys. Rev A \textbf{72}, 032108 (2005)
\bibitem{afanasev01} G. N. Afanasiev
Phys. Part. Nucl. \textbf{21}, 172 (1990).
\bibitem{avoid} J.-W. Ryu et al 
Phys. Rev. A \textbf{79}, 053858 (2009).
%
\bibitem{GS51} S. Granger and R.D. Spencer, Phys. Rev. \textbf{83}, 460 (1951).
\bibitem{Rassey58} A.J. Rassey,
 Phys. Rev. \textbf{109}, 949 (1958).
\bibitem{Arvieu87} Y. Ayant and R. Arvieu, J. Phys. A  \textbf{20}, 397 (1987). 
\bibitem{Arvieu93} F. Brut and R. Arvieu, J. Phys. A  \textbf{26}, 4749 (1993). 
\bibitem{Pashkevich69} V. V. Pashkevich and V. M. Strutinsky, 
Yad. Fiz. USSR \textbf{9}, 56 (1969).
\bibitem{P69ashkevich} J. Damgaard et al,
Nucl. Phys. A \textbf{135}, 432 (1969).
\bibitem{Greiner72} J. Maruhn and W. Greiner, Z. Physik \textbf{251}, 431 (1972).
\bibitem{Poenaru08} D.N. Poenaru et al, 
Phys. Lett. A \textbf{372}, 5448 (2008). 
\bibitem{Hofmann74} H. Hofmann, Nucl. Phys. A \textbf{224}, 116 (1974).
\bibitem{BuckPilt} B. Buck and A. A. Pilt,
Nucl. Phys A \textbf{80}, 133 (1977). 
\bibitem{Furman} S. G. Kadmensky, V. I. Furman,
 \emph{ Alpha decay and related nuclear Reactions} (Moscow, 1985).
\bibitem{cpc99} K. Hagino et al, 
Comput. Phys. Commun. 123, 143--152 (1999)
\bibitem{Zagrebaev04} V. I. Zagrebaev and V. V. Samarin, Phys. At. Nucl. \textbf{67}, 1462 (2004).
\bibitem{Zagrebaev07} V. I. Zagrebaev et.al., Phys. Part. Nucl. \textbf{38}, 469 (2007).
%
\bibitem{put}
L.V. Kantorovich and V.I. Krylov,  \emph{Approximate Methods of Higher Analysis}
(Wiley, NY, 1964).
\bibitem{CADE09} A.A. Gusev et al, Math. Comp. in Simulation  (2011) (accepted);
arXiv:1005.2089
\bibitem{BhK58} M. Born and  X. Huang,
 \emph{Dynamical Theory of Crystal Lattices}
(The Clarendon, Oxford, 1954).
\bibitem{CASC07} O. Chuluunbaatar et al, Lect. Notes Comp. Sci \textbf{4770},  118 (2007).
\bibitem{CASC09} S.I. Vinitsky et al, Lect. Notes Comp. Sci \textbf{5743}, 334 (2009).
\bibitem{CASC10} A.A. Gusev et al, Lect. Notes Comp. Sci. \textbf{6244}, 106 (2010).
%
\bibitem{ODPEVP} O. Chuluunbaatar et al, 
Comput. Phys. Commun.  \textbf{180}, 1358 (2009).
\bibitem{kantbp}  O. Chuluunbaatar et al, 
 Comput. Phys. Commun. \textbf{177}, 649  (2007).
%
\bibitem{Demkov1} Yu.N. Demkov
 JETP  \textbf{36}, 88-92 (1959).
\bibitem{Demkov} Yu.N. Demkov
JETP  \textbf{44}, 2007-2010 (1963).
\bibitem{Ilkaeva} L.A. Il'kaeva Vestnik LGU, 22, 56-63 (1963).
\bibitem{Erikson}  H.E. Erikson and E.L. Hill, Phys.Rev. \textbf{75}, 29 (1949).
\bibitem{MPST85} L.G. Mardoyan et al,
Preprint JINR, P2-85-139, Dubna, 1985.
\bibitem{stigun} M. Abramowitz and I.A. Stegun, \emph{Handbook of Mathematical
Functions} (Dover, New York,  1965).
%
\bibitem{MPST06} L.G. Mardoyan et al
\emph{Quantum systems with hidden symmetry}
(Fizmatlit, Moscow, 2006).
\bibitem{KKWP02} E. G. Kalnins et al
J. Math. Phys.  \textbf{43}, 3592 (2002). 
\bibitem{CurantGilbert} R. Courant and D. Hilbert, \emph{Methods of Mathematical
Physics. V. 1}  (Wiley, New York, 1989).
\bibitem{Harper} P.G. Harper,
Proc. Phys. Soc. A \textbf{68}, 874 (1955).
\bibitem{Akishin97} P.G. Akishin, F. Bosco, and S.I. Vinitsky, Comput. Math. Appl. \textbf{34}, 613 (1997).
\bibitem{Bayfield} J. E. Bayfield, \textit{Quantum Evolution
An Introduction to Time-Dependent
Quantum Mechanics} (John Wiley \& Sons, Inc., New York, 1999), p. 207.
\bibitem{BrunoCrespi1993} B. Crespi, G. Perez, and S.-J. Chang,
Phys.Rev. E \textbf{47}, 986 (1993).
\bibitem{Efros1982} Al.L. Efros, A.L. Efros, Sov. Phys. Semicond. \textbf{16}, 772  (1982).
\bibitem{LS1958} I.M. Lifshits and V.V. Slezov, Sov. Phys. JETF. \textbf{35}, 479  (1958).
\end{thebibliography}
\end{document}